\documentclass[10pt,epsfig,graphics,mathbbm]{article}

\usepackage{amsmath,amsfonts,amssymb,graphics,graphicx,epsfig,color,times,bbm}
\usepackage{color}

\newcommand{\me}{\mathrm{e}}
\newcommand{\mi}{\mathrm{i}}
\newcommand{\md}{\mathrm{d}}

\newcommand{\cSecondMomentsAmp}{c_1}
\newcommand{\cSecondMomentsMu}{\mu_1}
\newcommand{\cSecondMomentsEps}{\epsilon_1}

\newcommand{\cVariancesFP}{c_2}
\newcommand{\cFourthMomentsAmp}{c_3}
\newcommand{\cFourthMomentsEps}{\epsilon_2}
\newcommand{ \cFP}{c_4}

\newcommand{ \cCLEta}{\mu_2}
\newcommand{ \cCL}{c_5}

\newcommand{\cc}{\mathbbm{C}}
\newcommand{\zz}{\mathbbm{Z}}
\newcommand{\nn}{\mathbbm{N}}
\newcommand{\rr}{\mathbbm{R}}
\newcommand{\id}{\mathbbm{1}}
\renewcommand{\vec}[1]{\text{\boldmath$#1$}}

\newtheorem{theorem}{Theorem}
\newtheorem{lemma}{Lemma}
\newtheorem{corollary}{Corollary}

\newtheorem{assumption}{Assumption}

\begin{document}

\title{\Large\bf A quantum central limit theorem for non-equilibrium systems: Exact local relaxation of correlated states}

\author{\large
M.\ Cramer$^{1,2,3}$ and J.\ Eisert$^{3,4,5}$ }

\date{}
\maketitle

\vspace*{-.6cm}

\centerline{\it\footnotesize  1 Institut f\"ur Theoretische Physik, 
Albert-Einstein Allee 11, Universit\"at Ulm, D-89069 Ulm, Germany}

\centerline{\it\footnotesize  
2 QOLS, Blackett Laboratory, 
Imperial College London,
Prince Consort Road, London SW7 2BW, UK}

\centerline{\it\footnotesize  
3 Institute for Mathematical Sciences, Imperial College London,
Prince's Gardens, London SW7 2PE, UK}

\centerline{\it\footnotesize  
4 Institute for Physics and Astronomy, University of Potsdam, Karl-Liebknecht-Str., 14476 Potsdam, Germany}

\centerline{\it\footnotesize  
5 Institute for Advanced Study Berlin, Wallotstr., 14193 Berlin, Germany}

\maketitle
\begin{abstract}
We prove that quantum many-body systems on a one-dimensional lattice locally
relax to Gaussian states under non-equilibrium dynamics generated
by a bosonic quadratic Hamiltonian. This is true for a large class of initial states---pure or 
mixed---which have to satisfy merely weak conditions concerning the decay of correlations. 
The considered setting is a proven instance of a situation where
dynamically evolving closed quantum systems locally appear as if they had truly relaxed,
to maximum entropy states for fixed second moments. This furthers the 
understanding of relaxation in suddenly quenched quantum many-body systems.
The proof features a non-commutative central limit theorem for non-i.i.d.\ 
random variables, showing convergence to Gaussian characteristic functions, giving rise to  
trace-norm closeness. We briefly relate our findings to ideas of typicality and concentration of measure.
\end{abstract}

\tableofcontents

\section{Introduction}

In what sense can closed local many-body systems in a non-equilibrium situation
relax to an apparent equilibrium? Instances of that question have a long tradition in the literature.
This apparent contradiction of having entropy preserved in any closed system and 
at the same time arrive at an equilibrated situation for long times can yet be resolved
by acknowledging that merely {\it some} observables, and in particular entire 
subsystems, can well appear as if they had truly relaxed, even in closed systems. 
In local observations, such a composite system would then look entirely 
equilibrated. 

Early work formulating a quite similar intuition, albeit not in 
quantum lattice systems, was concerned with fermions freely moving in space 
\cite{Robinson}. The seminal work in Ref.\ \cite{Spohn,Spohn2} 
rigorously developed such an intuition 
for classical harmonic systems on cubic lattices. Refs.\ \cite{Lieb,Tegmark,Barthel} consider that 
question in instances of a bosonic or fermionic 
fully Gaussian setting, where initial states are Gaussian, 
and the evolution is governed by a quadratic Hamiltonian. Ref.\ \cite{simple} then 
arrives at a true local relaxation theorem for subsystems where initial states are not 
taken to be Gaussian: Yet, in the course of the non-equilibrium
dynamics of the system, locally, the states become Gaussian, so 
maximum entropy states for given second moments. This is true without 
having to invoke a time average. In such lattice models, the intuition is that non-equilibrium 
generates local excitations at each site that then travel ballistically through the lattice, 
resulting in a mixing at each site with excitations from further and further 
separated sites coming in \cite{simple,Calabrese}.

Some of the revived significant recent interest in equilibration and relaxation
in closed quantum many-body systems, needless to say, has been sparked off by novel
experiments with cold atoms in optical lattices or on 
low-dimensional structures. For theory work relating to those experiments,
see, e.g., Refs.\ \cite{Kollath,Flesch,Demler,RigolNew,Kehrein,Sachdev}
and references therein. This development is 
quite intriguing, as questions of an apparent local relaxation in non-equilibrium
dynamics can be measured and probed under very precisely tunable laboratory conditions.

In this work, we present a fully rigorous relaxation theorem, generalizing the 
findings of Ref.\ \cite{simple}, for one-dimensional quantum lattice systems. Physically speaking, the 
results presented here can be viewed as a relaxation theorem
forming a ``theoretical laboratory'', describing an idealized situation of non-equilibrium 
dynamics in the Bose-Hubbard setting: The system being initially held in some ground or thermal state
of the full Bose-Hubbard-Hamiltonian, and then suddenly switched or ``quenched'' to a 
parameter regime of strong hopping. 

It is shown that for general pure or mixed initial states satisfying
quite weak assumptions concerning the clustering of correlations---properties that
are expected to hold in particular for ground states of local many-body models---, 
locally, the system will relax to a Gaussian state under the dynamics
generated by a quadratic Hamiltonian. That is to say, we prove the ``local relaxation
conjecture'' for one-dimensional clustering states under quadratic Hamiltonians on a ring of $L$ sites. To wit, for an initial state $\hat{\varrho}_0$ that is evolved in time under a quadratic Hamiltonian and a  given finite subset $\mathcal{S}$ of the chain, we show that for every $\epsilon>0$ and  sufficiently large $L$ there is a relaxation time $t_{\text{relax}}$, and a recurrence time $t_{\text{rec}}$ (which grows unboundedly with $L$), such that the reduction of the time-evolved state to $\mathcal{S}$, $\hat{\varrho}_{\mathcal{S}}(t)$, fulfils 
\begin{equation}
\|\hat{\varrho}_{\mathcal{S}}(t)-\hat{\varrho}_G(t)\|_{\text{tr}}\le \epsilon\;\text{ for all }t_{\text{relax}}\le t\le t_{\text{rec}},
\end{equation}
in particular,
\begin{equation}
\lim_{t\rightarrow\infty}\lim_{L\rightarrow\infty}\|\hat{\varrho}_{\mathcal{S}}(t)-\hat{\varrho}_G(t)\|_{\text{tr}}=0.
\end{equation}
The state $\hat{\varrho}_G(t)$ is a Gaussian state with the same second moments as $\hat{\varrho}_{\mathcal{S}}(t)$. The proof features the explicit forms of
$t_{\text{relax}}$ and $t_{\text{rec}}$ in terms of the coupling parameters in the Hamiltonian and the decay behaviour of correlations of the initial state. 

The key ingredients in the proof are a quantum Lindeberg central limit theorem,
a notion of locality of dynamics as is manifest by a Lieb-Robinson bound, and
a Bernstein-Spohn-blocking argument reminiscent of the classical situation.
Essentially, we will show that for a given time excitations arising from
far away sites will hardly influence the dynamics at a subset of sites, whereas
the close sites give rise to a mixing such that a Lindeberg central limit theorem
can be invoked. In several ways, our quantum treatment reminds of the classical
situation presented in Ref.\ \cite{Spohn,Spohn2}, with the blocking argument being 
essentially identical, with significant differences in many other aspects.
In this work, we focus on the important case of a 
one-dimensional setting and allow for assumptions on the
initial state that can be relaxed, in order to render the argument
simpler than a fully general argument \cite{general}.

After rigorously proving convergence of characteristic functions to Gaussian functions in phase
space, we prove closeness in trace-norm of the respective reduced states.
In this way, we derive in the trace-norm topology a closeness to maximum entropy
states under the second moments. In some ways, the arguments reminds of the
situation of discrete time in convergence to stationary states in
quantum cellular automata \cite{QCA}. We also finally briefly discuss the findings in the
light of recent results concerning concentration of measure and notions
of typicality \cite{Goldstein,Popescu,Reimann}.

\section{Preliminaries}

In this section, we are collecting preliminaries that are being used in the proof. We also specify the 
model under consideration here and will specify the assumptions on the initial states in the subsequent
section. Note that the assumptions on such initial states are very mild and do by no means only
include pure or Gaussian states; merely the model Hamiltonian as such is 
taken to be a specific nearest-neighbor Hamiltonian.

\subsection{System and dynamical setting}

We consider a chain $\mathcal{L}=\{1,\dots, L\}$ equipped with periodic boundary conditions.
Each lattice site is associated with a bosonic canonical degree of freedom
equipped with the usual symplectic form.  
The distance between lattice sites $i,j\in\mathcal{L}$ is due to the periodic boundary conditions given by
\begin{equation}
d_{i,j}=\min\left\{|i-j|,|i-j+L|,|i-j-L|\right\}.
\end{equation}
For future purposes, we also define distances between subsets of sites. 
For $\mathcal{A},\mathcal{B}\subset \mathcal{L}$ we define their distance as
\begin{equation}
d_{\mathcal{A},\mathcal{B}}=\min_{i\in \mathcal{A},j\in \mathcal{B}}d_{i,j}.
\end{equation}

Starting from an initial quantum state (neither restricted to be a  pure state  nor a Gaussian state) $\hat{\varrho}_0$ on $\mathcal{L}$, we are concerned with its time evolution, 
\begin{equation}
\hat{\varrho}(t)=\me^{-\mi\hat{H}t}\hat{\varrho}_0\me^{\mi\hat{H}t},
\end{equation}
under the Hamiltonian
\begin{equation}
\hat{H}=\frac{1}{2}\sum_{i,j\in \mathcal{L}}\left(\hat{b}_i^\dagger A_{i,j}\hat{b}_j+\hat{b}_i A_{i,j}\hat{b}_j^\dagger
\right).
\end{equation}
$\hat{b}_i$ denotes a bosonic operator at site $i$.
We take the Hamiltonian to be defined by 
\begin{equation}
	A_{i,j}=-J\delta_{d_{i,j},1}.
\end{equation}
We will work in units in which  $J=1$, $\hbar=1$. This is the Hamiltonian
of a harmonic chain, in particular approximating the 
Hamiltonian of a Bose-Hubbard model in the
limit of a hopping that is dominant compared to the interaction.
Note that local relaxation may be shown for general quadratic Hamiltonians with 
sufficiently local couplings (exponentially decaying in the coupling distance) on (not necessarily cubic) 
lattices of arbitrary spatial dimension \cite{general}. The assumption that $\hat{H}$ is quadratic, 
is essential, however, for the validity of the argument. Yet, the physical intuition that local
relaxation is due to a mixing of excitations that essentially travel ballistically through the lattice
should be expected to be valid more generally. In this sense, the present proof constitutes 
the mentioned ``theoretical laboratory'' in which this mechanism can be studied in great clarity.

In particular, we will consider {\it local properties} of $\hat{\varrho}(t)$, and will see that in a sense
this reduced state will eventually relax in a sense yet to be specified. To this end let 
$\mathcal{S}\subset \mathcal{L}$ be a subset of sites of the lattice and define
the reduced state associated with the degrees of freedom of this subset as
\begin{equation}
\hat{\varrho}_\mathcal{S}(t)=\text{tr}_{\mathcal{L}\backslash \mathcal{S}}[\hat{\varrho}(t)].
\end{equation}
We write its characteristic function $\chi_{\hat{\varrho}_\mathcal{S}(t)}:\cc^{|\mathcal{S}|}\rightarrow\cc$ as
the expectation value of the Weyl operator, i.e., as
\begin{equation}
\chi_{\hat{\varrho}_\mathcal{S}(t)}(\vec{\beta})=\text{tr}_\mathcal{S}[\hat{\varrho}_\mathcal{S}(t)\hat{D}(\vec{\beta})],\;\;\;
\hat{D}(\vec{\beta})=\prod_{i\in \mathcal{S}}\me^{\beta_i\hat{b}_i^\dagger-\beta_i^*\hat{b}_i},
\end{equation}
as a complex-valued phase space function uniquely defining the state. States the characteristic function of
which are a Gaussian in phase space are referred to as being quasi-free or Gaussian states.

One finds, after solving Heisenberg's equation of motion and using the cyclic invariance of the trace,
for the time evolution of the characteristic function of the reduced state $\hat{\varrho}_\mathcal{S}(t)$
\begin{equation}
\chi_{\hat{\varrho}_\mathcal{S}(t)}(\vec{\beta})=\text{tr}_\mathcal{L}[\hat{\varrho}_0
\prod_{i\in \mathcal{L}}\me^{\alpha_i(t,\vec{\beta})\hat{b}_i^\dagger-\alpha^*_i(t,\vec{\beta})\hat{b}_i}
]=:
\text{tr}_\mathcal{L}[\hat{\varrho}_0
\hat{D}(\vec{\alpha}(t,\vec{\beta}))],
\end{equation}
where $\vec{\alpha}\in\cc^{|\mathcal{L}|}$,
\begin{equation}
\label{alpha}
\alpha_i=\sum_{j\in \mathcal{S}}\beta_jC_{j,i}^*(t),\;\;\;C=\me^{-\mi tA},
\end{equation}
the latter to be understood as a matrix exponent of the matrix $A$ collecting the coupling coefficients.
Note that this is exactly the setting of a sudden quench: One considers a ground or thermal state of some Hamiltonian (or any initial state) and the abrupt switch to the Hamiltonian $\hat H$, and follows the free time evolution under this new Hamiltonian. Note that the weak assumptions on the 
decay of correlations also allows for initial ground states of critical models.

\subsection{Moments}
For operators $\hat{O}$ defined on the Hilbert space associated with the lattice model, 
we write expectation values as
\begin{equation}
\langle\hat{O}\rangle=\text{tr}[\hat{\varrho}_0\hat{O}].
\end{equation}
For vectors of phase space coordinates
$\vec{\alpha},\vec{\alpha}^\prime\in\cc^{|\mathcal{L}|}$ we define the anti-hermitian operator
\begin{equation}\label{b}
\hat{b}(\vec{\alpha})=\sum_{i\in\mathcal{L}}\left(\alpha_i\hat{b}_i^\dagger-\alpha^*_i\hat{b}_i\right),
\end{equation}
and instances of second and fourth moments of $\hat{b}(\vec{\alpha})$ as
\begin{equation}
\sigma(\vec{\alpha})=\sigma(\vec{\alpha},\vec{\alpha}),\;\;\;
\sigma(\vec{\alpha},\vec{\alpha}^\prime)
=
\langle\hat{b}(\vec{\alpha})\hat{b}(\vec{\alpha}^\prime)\rangle,\;\;\;
f(\vec{\alpha})=\bigl\langle\bigl[\hat{b}(\vec{\alpha})\bigr]^4\bigr\rangle.
\end{equation}
For $\mathcal{A}\subset\mathcal{L}$ we define 
the vector $\vec{\alpha}_{\mathcal{A}}\in\cc^L$ by 
\begin{equation}
(\vec{\alpha}_{\mathcal{A}})_i=\begin{cases}
\alpha_i& \text{if }i\in\mathcal{A},\\
0& \text{otherwise,}
\end{cases}
\end{equation}
as the indicator function and write
\begin{equation}
\sigma_{\mathcal{A},\mathcal{B}}=\sigma(\vec{\alpha}_{\mathcal{A}},\vec{\alpha}_{\mathcal{B}}),\;\;\;
\sigma_{\mathcal{A}}=\sigma_{\mathcal{A},\mathcal{A}},\;\;\;
f_{\mathcal{A}}=f(\vec{\alpha}_{\mathcal{A}}),
\end{equation}
for certain second moments and functions of indicator functions.

\section{Assumptions on initial state and locality of dynamics}

In this section, we state the assumptions on 
the considered initial states and highlight the role of Lieb-Robinson 
bounds in the argument. 

\subsection{Assumptions on the initial state}

The assumptions on the initial state essentially express some degree of
clustering or decay of correlations. This clustering can be assumed to be 
quite weak, and even a slow algebraic decay of correlations is allowed
for. We will also include some natural conditions on the initial state that 
significantly simplify the argument. In particular, this includes an assumption
of the initial state to commute with the total particle number operator. In any
bosonic system involving massive particles, this natural requirement 
will always be satisfied.

More specifically, we assume that the initial state is 
such that\footnote{This simplifies the calculations significantly as expectation values of products involving unequal numbers of annihilation and creation operators vanish, is, however, not necessary to show local relaxation \cite{general}.}
\begin{equation}
\left[\hat{\varrho}_0,\sum_{i\in\mathcal{L}}\hat{b}_i^\dagger\hat{b}_i\right]=0.
\end{equation}
Then, for all $\vec{\alpha}\in\cc^L$ and all $\mathcal{A},\mathcal{B}\subset\mathcal{L}$ we have 
$\langle\hat{b}(\vec{\alpha})\rangle=0$, and
\begin{equation}
\label{moments}
\begin{split}
\sigma_{\mathcal{A},\mathcal{B}}&=\sigma(\vec{\alpha}_{\mathcal{A}},\vec{\alpha}_{\mathcal{B}})
=-
\sum_{i\in\mathcal{A}}\sum_{j\in\mathcal{B}}\alpha_i\alpha^*_j\left(\langle\hat{b}_i^\dagger\hat{b}_j\rangle+\langle\hat{b}_j\hat{b}_i^\dagger\rangle\right),\\
f_{\mathcal{A}}&=f(\vec{\alpha}_{\mathcal{A}})=
\sum_{i,j,k,l\in\mathcal{A}}\alpha_i\alpha_j\alpha^*_k\alpha^*_l
\Bigl(\langle\hat{b}_i^\dagger\hat{b}_j^\dagger\hat{b}_k
\hat{b}_l \rangle+\langle\hat{b}_i^\dagger\hat{b}_k\hat{b}_j^\dagger\hat{b}_l \rangle+\langle\hat{b}_i^\dagger\hat{b}_k\hat{b}_l\hat{b}_j^\dagger \rangle
\\
&\hspace{4cm}+\langle\hat{b}_k\hat{b}_j^\dagger\hat{b}_i^\dagger\hat{b}_l \rangle+\langle\hat{b}_k\hat{b}_j^\dagger\hat{b}_l\hat{b}_i^\dagger \rangle
+\langle\hat{b}_k\hat{b}_l\hat{b}_i^\dagger
\hat{b}_j^\dagger \rangle
\Bigr)
 .
\end{split}
\end{equation}
The first assumption concerns two-point correlations.

\begin{assumption}[Two-point correlations] \label{assumption:twoPoint} Let the initial state $\hat{\varrho}_0$ be such that there exist absolute constants $\cSecondMomentsAmp,\cSecondMomentsMu,\cSecondMomentsEps>0$ such that
\begin{equation}
\left|\langle\hat{b}^\dagger_i\hat{b}_j\rangle\right|
\le
\frac{\cSecondMomentsAmp}{\left[1+d_{i,j}\right]^{2+\cSecondMomentsMu+\cSecondMomentsEps}}
\end{equation}
for all $i,j$.
\end{assumption}
This is a natural assumption on the initial state in a way that does not depend on the system size.
This assumption may be 
relaxed to decay stronger than the spatial 
dimension of the lattice (i.e., in the setting at hand the two may be replaced by a one), resulting, 
however, in a much more involved proof \cite{general}. We note a first immediate 
consequence of Assumption \ref{assumption:twoPoint} that will be used in the main theorem. We also encounter the recurrence time: the system has to be sufficiently large for the bounds on the second moments to hold.

\begin{lemma}[Variances] \label{lemma:variances} Under Assumption \ref{assumption:twoPoint} and for 
$L\ge |t|^{7/6}\ge 1$ one has the following bounds.\footnote{We use the following vector norms for $\vec{\beta}\in\cc^{|\mathcal{S}|}$: 
$\|\vec{\beta}\|_1=\sum_{i\in\mathcal{S}}|\beta_i|$, $\|\vec{\beta}\|_2=\sqrt{\sum_{i\in\mathcal{S}}|\beta_i|^2}.$
}
\begin{equation}
|\sigma_{\mathcal{A},\mathcal{B}}|\le
\cVariancesFP\|\vec{\beta}\|^2_1  
\frac{\min\{|\mathcal{A}|,|\mathcal{B}|\}}{|t|^{2/3}\left[1+d_{\mathcal{A},\mathcal{B}}\right]^{1+\cSecondMomentsMu}}
\end{equation}
for all $\mathcal{A},\mathcal{B}\subset\mathcal{L}$,
\begin{equation}
\left|\sigma_{\mathcal{L}}-\sum_{i=1}^n\sigma_{\mathcal{A}_k}\right|\le|\sigma_{\mathcal{L}\backslash\mathcal{A}}|+2|\sigma_{\mathcal{L}\backslash\mathcal{A},\mathcal{A}}|
+\cVariancesFP\|\vec{\beta}\|^2_1 \sum_{i=1}^n
\tfrac{|\mathcal{A}_i|}{|t|^{2/3}\left[1+d_{\mathcal{A}\backslash \mathcal{A}_i,\mathcal{A}_i}\right]^{1+\cSecondMomentsMu}}
\end{equation}
for all $\mathcal{A}_i\subset\mathcal{L}$, $i=1,\dots,n$, $\mathcal{A}=\cup_i\mathcal{A}_i$, and
\begin{equation}
\begin{split}
|\sigma_{\mathcal{A},\mathcal{B}}|&\le
\cVariancesFP
\|\vec{\beta}\|^2_12^{-d_{\mathcal{A},\mathcal{S}}}
\end{split}
\end{equation}
for all $\mathcal{A},\mathcal{B}\subset \mathcal{L}$ such that $4\me |t|\le d_{\mathcal{A},\mathcal{S}}$.
Here,
\begin{equation}
\cVariancesFP=37^2(4\cSecondMomentsAmp+2)\zeta(1+\cSecondMomentsEps)
\end{equation}
and
 $\cSecondMomentsAmp$, $\cSecondMomentsEps$, and $\cSecondMomentsMu$ are as in Assumption \ref{assumption:twoPoint}.
\end{lemma}

\noindent The second assumption concerns four-point correlations.

\begin{assumption}[Four-point correlations] \label{assumption:fourPoint} Let the initial state 
$\hat{\rho}_0$ 
be such that there exist absolute constants $\cFourthMomentsAmp,\cFourthMomentsEps>0$ such that
\begin{equation}
\begin{split}
\Bigl|\Bigl\langle\hat{b}_i^\dagger\hat{b}_j^\dagger\hat{b}_k\hat{b}_l\Bigr\rangle\Bigr|
\le
\sum_{(r,s,t,u)\in P(i,j,k,l)}
\frac{\cFourthMomentsAmp}{([1+d_{r,s}][1+d_{t,u}])^{1+\cFourthMomentsEps}}
\end{split}
\end{equation}
for all $i,j,k,l$. Here, $P(i,j,k,l)$ contains all permutations of the indices $i,j,k,l$.
\end{assumption}

\begin{lemma}[Fourth moments] \label{lemma:fourthmoments} Under Assumptions \ref{assumption:twoPoint},\ref{assumption:fourPoint} and for  $L\ge |t|^{7/6}\ge 1$, one has
\begin{equation}
\sum_{i=1}^n|f_{\mathcal{A}_i}|
\le  
  \cFP
\|\vec{\beta}\|_2^2
\|\vec{\beta}\|^2_1
\tfrac{\max_{i}|\mathcal{A}_i|}{|t|^{2/3}}
\end{equation}
 for all $\mathcal{A}_i\subset\mathcal{L}$, $i=1,\dots,n$, with $\mathcal{A}_i\cap \mathcal{A}_j=\emptyset$ for $i\ne j$. Here,
 \begin{equation}
 \cFP=96(6\cFourthMomentsAmp+3(4\cSecondMomentsAmp+1))37^2\zeta^2(1+\min\{\cSecondMomentsEps,\cFourthMomentsEps\}),
 \end{equation}
 $\cSecondMomentsAmp$, $\cSecondMomentsEps$ are as in Assumption \ref{assumption:twoPoint} and
$\cFourthMomentsAmp$, $\cFourthMomentsEps$ as in Assumption \ref{assumption:fourPoint}.
\end{lemma}
We finally put a clustering condition on the initial state.
\begin{assumption}[Clustering] \label{Assumption:clustering} Let the initial state $\hat{\varrho}_0$ be such that there exist absolute constants $\cCL,\cCLEta>0$ such that
\begin{equation}
\left|
\langle\hat{D}(\vec{\alpha}_{\mathcal{A}})\hat{D}(\vec{\alpha}_{\mathcal{B}})\rangle
-\langle\hat{D}(\vec{\alpha}_{\mathcal{A}})\rangle\langle\hat{D}(\vec{\alpha}_{\mathcal{B}})\rangle\right|
\le \frac{\cCL}{[1+d_{\mathcal{A},\mathcal{B}}]^{1/2+\cCLEta}}
\end{equation}
for all $\mathcal{A},\mathcal{B}\subset\mathcal{L}$ and all $\mathcal{L}$.
\end{assumption}
Note that this condition may again be relaxed \cite{general}.

\subsection{Lieb-Robinson bounds and locality of dynamics}

Lieb-Robinson bounds are upper bounds to group velocities of excitations
travelling through a quantum lattice system. They define a causal cone of a 
local excitation, outside of which any influence of this excitation is exponentially
suppressed. As such, they provide an upper bound to the speed of any 
non-negligible information propagation
by time evolution of a quantum lattice model \cite{LR,LR2,LRH,Area}. Originally 
formulated for spin systems, in the present context, we 
need to invoke an instance that is suitable for the harmonic, 
infinite-dimensional individual constituents at hand \cite{LR3,LR4}. Subsequently we 
will formulate a useful related bound defining a causal cone for 
matrix entries of the propagator itself, which will be used in the main theorem. 

\begin{lemma}[Lieb-Robinson bound] \label{lemma:lieb_robinson} For all times, all 
$i,j\in\mathcal{L}$ and all $\mathcal{A}\subset\mathcal{L}$ with 
$4\me |t|\le d_{\mathcal{A},\mathcal{S}}$ one has
\begin{equation}
\begin{split}
|C_{i,j}|&\le \frac{|2t|^{d_{i,j}}}{d_{i,j}!},\;\;\;\sum_{i\in\mathcal{A}}|\alpha_i|\le
4
\|\vec{\beta}\|_12^{-d_{\mathcal{A},\mathcal{S}}}.
\end{split}
\end{equation}
For all $t,L$ such that $L\ge |t|^{7/6}\ge 1$ one has for all $i,j\in \mathcal{L}$ that
\begin{equation}
\label{all_entries}
\begin{split}
|C_{i,j}|&\le \tfrac{37}{|t|^{1/3}},\;\;\; |\alpha_i|\le\|\vec{\beta}\|_1\tfrac{37}{|t|^{1/3}}.
\end{split}
\end{equation}
\end{lemma}
This bound shows the exponential smallness of the matrix entries of $C=\me^{-\mi tA}$ in the distance $d_{i,j}$ away from the 
main diagonal at a given time. Also, Eqs.\ (\ref{all_entries}) signify bounds needed to prove convergence due to mixing within the causal cone for large times.
 The proof will be given in the appendix.

\section{Quantum central limit theorems}

Lindeberg central limit theorems are central limit theorems that 
do not rely on an identically distributed assumption concerning the considered
random variables. 
Here we formulate
a quantum version thereof, which will in the main theorem be extended to
a quantum version of a central limit theorem that in addition does not rely
on independent random variables.
We start by noting a number of useful facts
on quantum characteristic function in order to proceed to
formulate the quantum central limit theorem itself.

\subsection{Some facts about characteristic functions}

In this section, we will relate quantum characteristic functions to 
classical characteristic functions, by invoking Bochner's theorem.
For a similar correspondence of quantum to classical characteristic functions, 
see, e.g., Ref.\ \cite{Wolf}.
For a fixed vector of phase space coordinates 
$\vec{\alpha}\in\cc^{L}$ consider the function 
$\phi:\rr\rightarrow\cc$,
\begin{equation}
\phi(r)=\text{tr}[\hat{\varrho}_0\hat{D}(r\vec{\alpha})]=
\text{tr}[\hat{\varrho}_0\me^{r\hat{b}(\vec{\alpha})}]=
\langle\me^{r\hat{b}(\vec{\alpha})}\rangle.
\end{equation}
One finds $\phi^{(n)}(0)=0$ for odd $n\in\nn$ and
\begin{equation}
\phi^{(2)}(0)=\sigma(\vec{\alpha}),\;\;\;
\phi^{(4)}(0)=f(\vec{\alpha}),
\end{equation}
where $\phi^{(n)}$ denotes the $n$-th derivative of $\phi$.
Using the Cauchy-Schwarz inequality, 
one finds for even $m\in\nn$ the following chain of inequalities.
We suppress 
the argument $\vec{\alpha}$ of $\hat b$ defined as in Eq.\ (\ref{b}), 
use the fact that 
$\hat{b}^\dagger=-\hat{b}$ and write 
$\hat{\varrho}_0=\sum_n\varrho_n|\psi_n\rangle\langle \psi_n|$ for the spectral decomposition
of the state $\hat{\varrho}_0$. We hence find for the $m$-th derivative
\begin{equation}
\begin{split}
|\phi^{(m)}(r)|&\le
\sum_n\varrho_n\left|\langle \psi_n|  \hat{b}^{m/2}\me^{r\hat{b}}\hat{b}^{m/2} |\psi_n\rangle\right|\\
&\le
\sum_n\varrho_n\left|\langle \psi_n|  \hat{b}^m|\psi_n\rangle\right|
=\left|\sum_n\varrho_n\langle \psi_n|  \hat{b}^m|\psi_n\rangle\right|
=|\phi^{(m)}(0)|\\
&\le
\sum_n\varrho_n\left|\langle \psi_n|  (\hat{b}^\dagger)^m\hat{b}^m|\psi_n\rangle\right|^{1/2}\\
&\le
\left({\sum_n \varrho_n\langle \psi_n|\hat{b}^{2m}|\psi_n)}\right)^{1/2}
Ê=(\phi^{(2m)}(0))^{1/2}.
\end{split}
\end{equation}
Furthermore,
$\phi(0)=1$, and $\phi$ is positive semi-definite and continuous at the origin \cite{hudson}. It follows 
from Bochner's theorem that $\phi$ is a classical characteristic function of a classical random variable.
Hence, if $\phi^{(2)}(0)$ exists, $\phi$ is twice continuously differentiable on the entire real 
line, i.e., as $\phi^{(1)}(0)=0$, one has from Taylor's theorem
\begin{equation}
\left|\phi(1)-1\right|\le \int_0^1\md x\,|\phi^{(2)}(x)|(1-x)\le\frac{|\phi^{(2)}(0)|}{2}.
\end{equation}
Hence,
\begin{equation}
|\langle\hat{D}(\vec{\alpha})\rangle-1|\le\frac{|\sigma(\vec{\alpha})|}{2}
\end{equation}
for all $\vec{\alpha}\in\cc^L$ as $\vec{\alpha}$ was arbitrary.

If $\phi^{(4)}(0)$ exists, $\phi$ is four times continuously differentiable on the entire real line, i.e., one has from Taylor's theorem
\begin{equation}
\begin{split}
\left|\phi(1)-1-\frac{\phi^{(2)}(0)}{2}\right|
\le \int_0^1\md x\,|\phi^{(4)}(x)|\frac{(1-x)^3}{6}
\le\frac{|\phi^{(4)}(0)|}{24}.
\end{split}
\end{equation}

\subsection{A Lindeberg-type quantum central limit theorem}

We now turn to the actual quantum instance of a Lindeberg-type
central limit theorem.
Let $\mathcal{A}_1,\dots,\mathcal{A}_n\subset \mathcal{L}$ be mutually disjoint sets. These take the role of independent but not necessarily identically distributed random variables. Consider 
\begin{equation}
z_i=\phi_{i}(1)=\langle\exp[\hat{b}(\vec{\alpha}_{{\cal A}_i})]\rangle,\;\;\; i=1,\dots,n.
\end{equation}
If $|z_i-1|\le 1/2$ for all $i$, a complex logarithm is defined by the Mercator series. We have
\begin{equation}
\begin{split}
\left|\log(z_i)-\phi_i^{(2)}(0)/2\right|&\le
\left|\log(z_i)-(z_i-1)\right|+\left|z_i-1-\phi_i^{(2)}(0)/2\right|\\
&\le\left|\log(z_i)-(z_i-1)\right|+\frac{|\phi_i^{(4)}(0)|}{24},
\end{split}
\end{equation}
where
\begin{equation}
\begin{split}
\left|\log(z_i)-(z_i-1)\right|&=\left|\log(1+(z_i-1))-(z_i-1)\right|\le
\sum_{n=2}^\infty\frac{|z_i-1|^{n}}{n}\\
&=
|z_i-1|^2\sum_{n=0}^\infty\frac{|z_i-1|^{n}}{n+2}
\le |z_i-1|^2\sum_{n=0}^\infty\frac{|1/2|^{n}}{n+2}\\
&\le
|z_i-1|^2
\le
\frac{|\phi_i^{(2)}(0)|^2}{4}
\le
\frac{|\phi_i^{(4)}(0)|}{4}
.
\end{split}
\end{equation}
Hence,
\begin{equation}
\left|\sum_i\log(z_i)-\sum_i\frac{\sigma_{\mathcal{A}_i}}{2}\right|\le
\frac{7}{24}\sum_i|\phi_i^{(4)}(0)|.
\end{equation}
Now, let $x,y,z\in\rr$. We set out to show that $|\me^{x+\mi y}-\me^{z}|\le |x-z+\mi y|\me^{\max\{x,z\}}$. This holds for $x=z$, 
i.e., we may let $x\ne z$ w.l.o.g. Then
\begin{equation}
\me^{x+\mi y}-\me^{z}=\me^z(\me^{x-z+\mi y}-1)=\me^z(x-z+\mi y)\int_0^1\md t\,\me^{t(x-z+\mi y)},
\end{equation}
i.e., using the mean value theorem, there is a $c$ between $x,z$ such that
\begin{equation}
\begin{split}
|\me^{x+\mi y}-\me^{z}|&\le \me^z|x-z+\mi y|\int_0^1\md t\,\me^{t(x-z)}=|x-z+\mi y|\frac{\me^{x}-\me^z}{x-z}\\
&=|x-z+\mi y|\me^c
\le |x-z+\mi y|\me^{\max \{x,z\}}.
\end{split}
\end{equation}
Now, we recall that  $|z_i-1|\le 1/2$ and $|z_i|\le 1$, i.e., 
\begin{equation}
	\log |z_i|=-|\log |z_i||. 
\end{equation}
Hence, using the fact that $\sigma_{\mathcal{A}_i}=-|\sigma_{\mathcal{A}_i}|$, we find
\begin{equation}
\begin{split}
\left|\prod_iz_i-\me^{\sum_i\sigma_{\mathcal{A}_i}/2}\right|&=\left|\prod_i\me^{\log z_i}-\me^{-\sum_i|\sigma_{\mathcal{A}_i}|/2}\right|\\
&=\left|\me^{-\sum_i|\log |z_i||+\mi \sum_i\arg z_i}-\me^{-\sum_i|\sigma_{\mathcal{A}_i}|/2}\right|\\
&\le
\left|\sum_i\log z_i-\sum_i\frac{\sigma_{\mathcal{A}_i}}{2}\right|\me^{\max \{-\sum_i|\log |z_i||,-\sum_i|\sigma_{\mathcal{A}_i}|/2\}}\\
&\le
\left|\sum_i\log z_i-\sum_i\frac{\sigma_{\mathcal{A}_i}}{2}\right|,
\end{split}
\end{equation}
which yields the following theorem.

\begin{theorem}[Lindeberg quantum central limit theorem] 
Let $\mathcal{A}_i\subset \mathcal{L}$, $i=1,\dots, n$, be mutually disjoint sets, $|\sigma_{\mathcal{A}_i}|\le 1$, and let $f_{\mathcal{A}_i}<\infty$. Then
\begin{equation}
\left|\prod_i\langle\hat{D}(\vec{\alpha}_{\mathcal{A}_i})\rangle-\me^{\sum_i\sigma_{\mathcal{A}_i}/2}\right|\le
\frac{7}{24}\sum_{i=1}^n|f_{\mathcal{A}_i}|.
\end{equation}
\end{theorem}
A dynamical quantum central limit for initially uncorrelated states $\hat{\varrho}_0=\otimes_{i=1}^L \hat{\varrho}_i$ follows directly from this theorem. 
The subsequent technicalities are due to the fact that we allow for correlations in the initial state.

\section{Main theorem}

We now turn to stating the main theorem. We arrive at a statement showing local
convergence to a Gaussian state in trace-norm, under the weak assumptions on the initial
state specified above. 
We first introduce a variant of a Bernstein-Spohn blocking argument, and then
make use of the above quantum Lindeberg central limit theorem, to show convergence
to a Gaussian characteristic function in phase space. We then show that this implies
closeness of $\hat{\varrho}_{\mathcal{S}}(t)$ to a Gaussian state---so a maximum entropy state for given
second moments. 

\subsection{The Bernstein-Spohn blocking argument}

We divide---for a given time---the 
lattice into two parts: Into sites inside the causal cone, where mixing occurs, 
and sites outside the causal cone, 
the influence of which is exponentially suppressed. Inside the cone, we in turn split the 
region into several subsets, the role of which will soon become obvious.
This choice of a blocking is derived from the ``room and corridor" argument on classical lattice systems
presented in Ref.\ \cite{Spohn,Spohn2}.
To this end let 
\begin{equation}
a\ge b\ge 1 \text{ such that }n:=
\left\lfloor \frac{8\me |t|}{a+b}\right\rfloor>1.
\end{equation}
Then
\begin{equation}
8\me |t|\ge n(a+b)\ge 4\me |t|.
\end{equation}
Now define the set
\begin{equation}
\mathcal{T}=\left\{j\in L\,:\,d_{j,\mathcal{S}}\ge n(a+b) \right\}
\end{equation}
as the sites that are significantly outside the causal cone, compare Lemma \ref{lemma:lieb_robinson}. Define also the sets 
$\mathcal{A}=\bigcup_{i=1}^n\mathcal{A}_i$ and $\mathcal{B}=\bigcup_{i=1}^n\mathcal{B}_i$,
where
\begin{equation}
\begin{split}
\mathcal{A}_i&=\left\{j\in \mathcal{L}\,:\, (i-1)(a+b)\le d_{j,\mathcal{S}}< i(a+b)-b\right\},\\
\mathcal{B}_i&=\left\{j\in L\,:\, i(a+b)-b \le d_{j,\mathcal{S}}< i(a+b)\right\}.
\end{split}
\end{equation}
Then $\mathcal{L}=\mathcal{A}\cup \mathcal{B}\cup \mathcal{T}$---the entire lattice is covered by these sets---and 
$\mathcal{S}\subset \mathcal{A}_1$. What is more, 
$\mathcal{A}\cap \mathcal{B}=\mathcal{A}\cap \mathcal{T}=
\mathcal{T}\cap \mathcal{B}=\mathcal{A}_i\cap \mathcal{A}_j=\mathcal{B}_i\cap \mathcal{B}_j=\emptyset$ for $i\ne j$. 
The sets $\mathcal{A}$ and $\mathcal{B}$ cover hence the complement of  $\mathcal{T}$ in an 
alternating fashion, the first set $\mathcal{A}_1$ containing the subset $\mathcal{S}$ of interest for
which we aim at showing relaxation. The goal is now to choose $b$ sufficiently small (such that the influence of sites in $\mathcal{B}$ is small due to the large separation of the $\mathcal{B}_i$), but large enough such that the $\mathcal{A}_i$ are sufficiently far apart (and therefore approximately independent) and the Lindeberg central limit theorem yields closeness of $\chi_{\hat{\varrho}_{\mathcal{S}}(t)}$ to a Gaussian function in phase space. We will see that for sufficiently large $|t|$, this is achieved by the choice
\begin{equation}
a=\frac{|t|^{2/3}}{\log|t|},\;\;\; b=|t|^{(2-\cSecondMomentsMu/(1+\cSecondMomentsMu))/6},
\end{equation} 
for which one can prove the following Lemma.
\begin{lemma}[Correlations]
\label{lemma:bernstein} Let Assumptions \ref{assumption:twoPoint} and \ref{assumption:fourPoint} hold and let $t$, $L$ be such that 
\begin{equation}
L^{6/7}\ge |t|\ge 2,\;\;\;
\log |t|\le |t|^{1/3+\mu},
\end{equation}
where $\mu=\cSecondMomentsMu/(6(\cSecondMomentsMu+1))$.
Then, 
\begin{equation}
\frac{|t|^{2/3}}{\log |t|}=a\ge b=|t|^{1/3-\mu}\ge 1,\;\;\; \left\lfloor\frac{8\me |t|}{a+b}\right\rfloor =n>1,
\end{equation}
and
for $\mathcal{A}$, $\mathcal{B}$, $\mathcal{T}$ as above, one has 
\begin{equation}
\begin{split}
|\sigma_{\mathcal{T}}|,|\sigma_{\mathcal{B},\mathcal{T}}|,|\sigma_{\mathcal{A},\mathcal{T}}|&\le 
\cVariancesFP
\|\vec{\beta}\|^2_12^{-4\me |t|},\\
|\sigma_{\mathcal{B}}|,|\sigma_{\mathcal{A},\mathcal{B}}|&\le 
16\me\cVariancesFP |\partial\mathcal{S}| \|\vec{\beta}\|^2_1  
   \tfrac{\log|t|}{|t|^{\mu}}
,\\
\left|\sigma_{\mathcal{L}}-\sum_{i=1}^n\sigma_{\mathcal{A}_i}\right|&\le
\cVariancesFP\|\vec{\beta}\|^2_1\left(
\tfrac{2|\mathcal{S}|}{|t|^{2/3}}
+80\me|\partial\mathcal{S}|
 \tfrac{\log|t|}{|t|^{\mu}}\right),\\
\sum_{i=1}^n|f_{\mathcal{A}_i}|
&\le
\cFP
\|\vec{\beta}\|_2^2\|\vec{\beta}\|_1^2\left(
\tfrac{|\mathcal{S}|}{|t|^{2/3}}+\tfrac{2|\partial\mathcal{S}|}{\log |t|}\right).
  \end{split}
\end{equation}
\end{lemma}
The proof is again presented in the appendix. Here, the constant $\mu_1$ is defined as specified
in the assumptions on the initial state, $\cVariancesFP$, $\cFP$ as in the Lemmata following the assumptions, and
\begin{equation}
\partial\mathcal{S}=\left\{i\in\mathcal{S}\, :\, d_{i,\mathcal{L}\backslash\mathcal{S}}=1\right\}
\end{equation}
is the boundary of $\mathcal{S}$.

\subsection{The main theorem}

Let $\mathcal{A}$, $\mathcal{B}$, $\mathcal{T}$ as above and write $\hat{D}_{\mathcal{A}}=\hat{D}(\vec{\alpha}_{\mathcal{A}})$ and similarly for other the sets.
We have (we employ the triangle inequality three times)
\begin{equation}
\label{mainThing}
\begin{split}
\left|\chi_{\hat{\varrho}_{\mathcal{S}}(t)}(\vec{\beta})-\me^{\sigma_{\mathcal{L}}/2}\right|
&\le
\left|\langle\hat{D}_{\mathcal{L}}\rangle-\langle\hat{D}_{\mathcal{A}}\rangle\right|
+\left|\langle\hat{D}_{\mathcal{A}}\rangle-\prod_{i=1}^n\langle\hat{D}_{\mathcal{A}_i}\rangle\right|\\
&\hspace{0.5cm}+\left|\me^{\sum_{i=1}^n\sigma_{\mathcal{A}_i}/2}-\me^{\sigma_{\mathcal{L}}/2}\right|
+\left|\prod_{i=1}^n\langle\hat{D}_{\mathcal{A}_i}\rangle-\prod_{i=1}^n\me^{\sigma_{\mathcal{A}_i}/2}\right|.
\end{split}
\end{equation}
The first term is bounded by 
(we write again $\hat{\varrho}_0=\sum_n\varrho_n|\psi_n\rangle\langle \psi_n|$ for the spectral decomposition of 
$\hat{\varrho}_0$
and employ the Cauchy-Schwarz inequality)
\begin{equation}
\begin{split}
\left|\langle\hat{D}_{\mathcal{A}}(\hat{D}_{\mathcal{B}\cup\mathcal{T}}-\id)\rangle\right|&\le
\sum_n\varrho_n|\langle \psi_n|   \hat{D}_{\mathcal{A}}(\hat{D}_{\mathcal{B}\cup\mathcal{T}}-\id)   |\psi_n\rangle|\\
&\le
\sum_n\sqrt{\varrho_n}\left({\varrho_n\langle \psi_n|   
(\hat{D}_{\mathcal{B}\cup\mathcal{T}}-\id)^\dagger(\hat{D}_{\mathcal{B}\cup\mathcal{T}}-\id)   |\psi_n\rangle}\right)^{1/2}\\
&\le\left({\langle   (\hat{D}_{\mathcal{B}\cup\mathcal{T}}-\id)^\dagger(\hat{D}_{\mathcal{B}\cup\mathcal{T}}-\id)  \rangle}\right)^{1/2}\\
&=\left({2\Re[\langle   (\id-\hat{D}_{\mathcal{B}\cup\mathcal{T}}) \rangle]}\right)^{1/2}
\le{|\sigma_{\mathcal{B}\cup\mathcal{T}}|^{1/2}}\\
&\le
\left({|\sigma_{\mathcal{B}}|+|\sigma_{\mathcal{T}}|+2|\sigma_{\mathcal{B},\mathcal{T}}|}\right)^{1/2}.
\end{split}
\end{equation}
To bound the second term in Eq.\ (\ref{mainThing}) we let $I=\{m\in\nn\,:\,m\text{ even},1\le m\le n\}$, 
$J=\{1,\dots,n\}\backslash I$. Then, using the fact that $d_{\mathcal{A}_i,\mathcal{A}_j}\ge b$ for $i\ne j$, we find under Assumption \ref{Assumption:clustering}
\begin{equation}
\begin{split}
\left|\langle\prod_{i=1}^n\hat{D}_{\mathcal{A}_i}\rangle-\prod_{i=1}^n\langle\hat{D}_{\mathcal{A}_i}\rangle\right|
\le
\left|\langle\prod_{i\in I}\hat{D}_{\mathcal{A}_i}\prod_{j\in J}\hat{D}_{\mathcal{A}_j}\rangle-
\langle\prod_{i\in I}\hat{D}_{\mathcal{A}_i}\rangle\langle\prod_{j\in J}\hat{D}_{\mathcal{A}_j}\rangle
\right|\\
+
\left|\langle\prod_{i\in I}\hat{D}_{\mathcal{A}_i}\rangle\langle\prod_{j\in J}\hat{D}_{\mathcal{A}_j}\rangle-\prod_{i=1}^n\langle\hat{D}_{\mathcal{A}_i}\rangle
\right|\\
\le
\frac{\cCL}{b^{1/2+\cCLEta}}
+
\left|\langle\prod_{i\in I}\hat{D}_{\mathcal{A}_i}\rangle\langle\prod_{j\in J}\hat{D}_{\mathcal{A}_j}\rangle
-\langle\prod_{i\in I}\hat{D}_{\mathcal{A}_i}\rangle\prod_{j\in J}\langle\hat{D}_{\mathcal{A}_i}\rangle
\right|\\
+
\left|\langle\prod_{i\in I}\hat{D}_{\mathcal{A}_i}\rangle\prod_{j\in J}\langle\hat{D}_{\mathcal{A}_i}\rangle-\prod_{i=1}^n\langle\hat{D}_{\mathcal{A}_i}\rangle
\right|\\
\le
\frac{\cCL}{b^{1/2+\cCLEta}}
+
\left|\langle\prod_{j\in J}\hat{D}_{\mathcal{A}_j}\rangle
-\prod_{j\in J}\langle\hat{D}_{\mathcal{A}_j}\rangle
\right|
+
\left|\langle\prod_{i\in I}\hat{D}_{\mathcal{A}_i}\rangle-\prod_{i\in I}\langle\hat{D}_{\mathcal{A}_i}\rangle
\right|
\end{split}
\end{equation}
and, using the fact that $d_{\mathcal{A}_i,\mathcal{A}_j}\ge a$ for $i, j\in I$, $i\ne j$,
\begin{equation}
\begin{split}
\left|\langle\prod_{i\in I}\hat{D}_{\mathcal{A}_i}\rangle-\prod_{i\in I}\langle\hat{D}_{\mathcal{A}_i}\rangle
\right|\le
\left|\langle\prod_{i\in I}\hat{D}_{\mathcal{A}_i}\rangle-\langle\hat{D}_{\mathcal{A}_2}\rangle\langle\prod_{2\ne i\in I}\hat{D}_{\mathcal{A}_i}\rangle
\right|\\
+
\left|\langle\prod_{2\ne i\in I}\hat{D}_{\mathcal{A}_i}\rangle-\prod_{2\ne i\in I}\langle\hat{D}_{\mathcal{A}_i}\rangle
\right|\\
\le \frac{\cCL}{a^{1/2+\cCLEta}}+
\left|\langle\prod_{2\ne i\in I}\hat{D}_{\mathcal{A}_i}\rangle-\prod_{2\ne i\in I}\langle\hat{D}_{\mathcal{A}_i}\rangle
\right|\\
\le\cdots\le |I|\frac{\cCL}{a^{1/2+\cCLEta}}.
\end{split}
\end{equation}
This means that
\begin{equation}
\begin{split}
\left|\langle\prod_{i=1}^n\hat{D}_{\mathcal{A}_i}\rangle-\prod_{i=1}^n\langle\hat{D}_{\mathcal{A}_i}\rangle\right|
&\le
\frac{\cCL}{b^{1/2+\cCLEta}}+\frac{\cCL n}{a^{1/2+\cCLEta}}\\
&\le
\frac{\cCL}{|t|^{(1/12+\cCLEta/6)}}+8\cCL \me\frac{  (\log|t|)^{3/2+\cCLEta}}{|t|^{2\cCLEta/3}},
\end{split}
\end{equation}
which bounds the second term in Eq.\ (\ref{mainThing}).

Now, for $x,y\in\rr$ there is a $x_0$ between $x$ and $y$ such that
\begin{equation}
\me^x-\me^y=(x-y)\me^{x_0},\;\text{i.e.,}\; |\me^x-\me^y|=|x-y|\me^{x_0}\le |x-y|\me^{\max\{x,y\}}.
\end{equation}
Hence, for the third term in Eq.\ (\ref{mainThing}) we find (we recall that $\sigma_{\mathcal{L}},\sigma_{\mathcal{A}_i}\le 0$) 
\begin{equation}
\left|\me^{\sum_{i=1}^n\sigma_{\mathcal{A}_i}/2}-\me^{\sigma_{\mathcal{L}}/2}\right|\le
\left|\sigma_{\mathcal{L}}-\sum_{i=1}^n\sigma_{\mathcal{A}_i}\right|/2.
\end{equation}
If $|\sigma_{\mathcal{A}_i}|\le 1$, the last term in Eq.\ (\ref{mainThing}) may be bounded using the Lindeberg-type central limit theorem 
and we finally have
\begin{equation}
\begin{split}
\left|\chi_{\hat{\varrho}_{\mathcal{S}}(t)}(\vec{\beta})-\me^{\sigma_{\mathcal{L}}/2}\right|
&\le
\left({|\sigma_{\mathcal{B}}|+|\sigma_{\mathcal{T}}|+2|\sigma_{\mathcal{B},\mathcal{T}}|}\right)^{1/2}
\\
&\hspace{0.5cm}+\left|\sigma_{\mathcal{L}}-\sum_{i=1}^n\sigma_{\mathcal{A}_i}\right|/2
+\frac{7}{24}\sum_{i=1}^n|f_{\mathcal{A}_i}|\\
&\hspace{0.5cm}+\frac{\cCL}{|t|^{(1/12+\cCLEta/6)}}+8\cCL \me\frac{  (\log|t|)^{3/2+\cCLEta}}{|t|^{2\cCLEta/3}},
\end{split}
\end{equation}
where bounds on the moments may be read off 
from Lemma \ref{lemma:bernstein}. This yields the main theorem.

\begin{theorem}[Convergence of characteristic functions]\label{Main}
For all $\mathcal{L}=\{1,\dots, L\}\subset\nn$, all $\mathcal{S}\subset\mathcal{L}$, and every $\hat{\varrho}_0$ fulfilling Assumptions \ref{assumption:twoPoint}-\ref{Assumption:clustering}
there exists a $F_{\mathcal{S}}(\vec{\beta},t,L)>0$ such that
\begin{equation}
\left|\chi_{\hat{\varrho}_{\mathcal{S}}(t)}(\vec{\beta})-\me^{\sigma_{\mathcal{L}}/2}\right|\le F_{\mathcal{S}}(\vec{\beta},t,L).
\end{equation}
The function $F_{\mathcal{S}}:\cc^{|\mathcal{S}|}\times\rr\times\nn\rightarrow\rr$ has the following properties. There is a recurrence time $t_{\text{rec}}>0$, which increases polynomially with $L$, and a relaxation time $t_{\text{relax}}$ such that $F_{\mathcal{S}}$ is decreasing in time
for all $|t|\in [t_{\text{relax}},t_{\text{rec}}]$,  and all $\mathcal{S}\subset\{1,\dots, L\}$ with
\begin{equation}
\frac{|\partial\mathcal{S}|}{\log|t|},\frac{|\mathcal{S}|}{|t|^{2/3}}\;\substack{\longrightarrow \\ |t|\rightarrow\infty}\;0.
\end{equation}
If the latter holds and $|\mathcal{S}|$ does not depend on $L$, one has in particular
\begin{equation}
\lim_{|t|\rightarrow\infty}\lim_{L\rightarrow\infty}F_{\mathcal{S}}(\vec{\beta},t,L)=0.
\end{equation}
\end{theorem}

This theorem 
indeed shows convergence of characteristic functions of reduced 
states associated with some sublattice ${\cal S}$
to Gaussian characteristic functions in phase space.
Note that the size of this sublattice may even slowly grow in time; one would
still end up with a Gaussian characteristic function. Surely, in any finite system there
will be recurrences, but with increasing $L$, these recurrence times $t_{\text{rec}}$  can be made arbitrarily large.

The characteristic function
\begin{equation}
\chi_{\hat{\varrho}_G(t)}(\vec{\beta})=\me^{\sigma_{\mathcal{L}}/2}=\me^{-\vec{\beta}^\dagger\gamma(t)\vec{\beta}/2}
\end{equation}
corresponds to the Gaussian state $\hat{\varrho}_G(t)$ with second moments
\begin{equation}
[\gamma(t)]_{l,k}=\langle \hat{b}^\dagger_k(t)\hat{b}_l(t)\rangle
+
\langle  \hat{b}_l(t) \hat{b}_k^\dagger(t)\rangle
=[\me^{-\mi tA}\gamma(0)\me^{\mi tA}]_{l,k}
\end{equation}
and one has\footnote{Here, $\|\cdot\|$ denotes the spectral norm.}
\begin{equation}
\begin{split}
\text{tr}[\hat{\varrho}_G(t)\hat{b}_i^\dagger\hat{b}_i]&=\text{tr}[\hat{\varrho}_{\mathcal{S}}(t)\hat{b}_i^\dagger\hat{b}_i]\\
&=\frac{[\gamma(t)]_{i,i}-1}{2}\le\|\me^{-\mi tA}\gamma(0)\me^{\mi tA}\|=\|\gamma(0)\|.
\end{split}
\end{equation}
Note that, if the initial correlations are translationally invariant, $\gamma(0)$ and $A$ commute, i.e., the second moments are conserved, $\gamma(t)=\gamma(0)$.
By virtue of Lemma \ref{Close}, which we prove in the next section,
trace-norm convergence of states is also inherited from this main theorem. We state this
corollary for a sublattice ${\cal S}$ that is constant in time. 

\begin{corollary}[Convergence of reduced states to Gaussian states] \label{mainCorr} Let $|\mathcal{S}|$ and $\|\gamma(0)\|$ be bounded by an absolute  constant.
Under the assumptions of Theorem \ref{Main} and for any $\varepsilon>0$ there exists a recurrence time $t_{\text{rec}}>0$, which increases polynomially with $L$, and a relaxation time $t_{\text{relax}}$ such that
\begin{equation}
	\|\hat{\varrho}_{\cal S}(t)- \hat{\varrho}_G(t)\|_{\text{tr}} \leq \varepsilon,
\end{equation}
for all $|t|\in [t_{\text{relax}},t_{\text{rec}}]$,
where $\hat{\varrho}_G(t)$ is the Gaussian state with characteristic function $\me^{\sigma_{\mathcal{L}}/2}$.
\end{corollary}
This means that in the sense of trace-norm closeness, local reduced states become Gaussian
---and hence maximum entropy---states.

\section{Convergence of quantum states}

In this section, we will show that pointwise convergence of characteristic functions is inherited
by trace-norm convergence on the level of quantum states. This argument
generalizes the one of Ref.\  \cite{hudson}---showing weak convergence for 
states from pointwise convergence of characteristic functions---and of Ref.\ \cite{trace} -- relating
trace-norm and weak convergence. The central idea
is to consider two sets: On the one hand, this is the subset of 
phase space that supports, in a sense yet to be made precise,
``most'' of the characteristic function. On the other hand, this is 
the subspace of Hilbert space with the property that the projection of the state onto 
it is a positive operator that eventually has almost unit trace. One can then bound the error
made when neglecting the complement of these sets and in turn relate the relevant sets in phase and state space.
Inside these relevant sets, one can relate weak convergence to pointwise convergence of 
characteristic functions and use the equivalence of norms. Here we present an argument
going beyond the results in Refs.\ \cite{hudson,trace}
and use a gentle measurement Lemma of Ref.\ \cite{gentle}. 

\subsection{Preliminaries}

To prepare the argument,
consider for a single mode the upper bound to matrix elements of Weyl operators
\begin{equation}
\begin{split}
\left|\langle n|\hat{D}(\alpha)|m\rangle\right|&\le
\me^{-|\alpha|^2/2}|\alpha|^{|n-m|}
\left({\frac{\min\{m,n\}!}{\max\{m,n\}!}}
\right)^{1/2}
\left|L_{\min\{m,n\}}^{|n-m|}(|\alpha|^2)\right|\\
&=:\me^{-|\alpha|^2/2}p_{n,m}(|\alpha|^2)
,
\end{split}
\end{equation}
where $\alpha\in \cc$ and
$L_n^k$ are associated Laguerre polynomials, i.e., $p_{n,m}(|\alpha|^2)$ is a polynomial of degree 
$\min\{m,n\}+|n-m|/2=(n+m)/2$ in $|\alpha|^2$. 
Here $\{|n\rangle\}_{n=0,1,\dots}$ is the standard number basis of the Hilbert space
associated with a single mode. We can make use of this in the 
situation of having $N$ modes: Define for some $T>0$ 
\begin{equation}
	\bar{\mathcal{M}}:=\left\{\vec{\alpha}\in\cc^N\, :\, \|\vec{\alpha}\|_2^2>T\right\}, 
\end{equation}
as the region far away in $2$-norm from the origin of phase space
and $\mathcal{M}:=\cc^N\backslash\bar{\mathcal{M}}$ as its complement. 
For a continuous function $f:\cc^N\rightarrow\cc$ with 
$|f(\vec{\alpha})|\le 2$ consider the bound (we write $p_{\vec{n},\vec{m}}(\vec{\alpha})=\prod_ip_{n_i,m_i}(|\alpha_i|^2)$ and $|\vec{n}\rangle=|n_1\cdots n_N\rangle$)
\begin{equation}
\begin{split}
\left|\int_{\cc^{N}}\!\!\!\!\!\!\md\vec{\alpha}\,f(\vec{\alpha})
\langle\vec{n}|\hat{D}(\vec{\alpha})
|\vec{m}\rangle\right|
&\le
\int_{\bar{\mathcal{M}}}\!\!\!\!\md\vec{\alpha}\,\left|f(\vec{\alpha})\right|
\left|\langle\vec{n}|\hat{D}(\vec{\alpha})
|\vec{m}\rangle\right|\\
&\hspace{1cm}+
\int_{\mathcal{M}}\!\!\!\!\md\vec{\alpha}\,\left|f(\vec{\alpha})\right|
\left|\langle\vec{n}|\hat{D}(\vec{\alpha})
|\vec{m}\rangle\right|\\
&\le
2\int_{\bar{\mathcal{M}}}\!\!\!\!\md\vec{\alpha}\,
\left|\langle\vec{n}|\hat{D}(\vec{\alpha})
|\vec{m}\rangle\right|
+
|\mathcal{M}|\max_{\vec{\alpha}\in\mathcal{M}}|f(\vec{\alpha})|\\
&\le
2\int_{\bar{\mathcal{M}}}\!\!\!\!\md\vec{\alpha}\,
\me^{-\|\vec{\alpha}\|_2^2/2}p_{\vec{n},\vec{m}}(\vec{\alpha})
+
|\mathcal{M}|\max_{\vec{\alpha}\in\mathcal{M}}|f(\vec{\alpha})|
\\
&\le
2\me^{-T/4}\int_{\cc^N}\!\!\!\!\!\!\md\vec{\alpha}\,
\me^{-\|\vec{\alpha}\|_2^2/4}p_{\vec{n},\vec{m}}(\vec{\alpha})
+
|\mathcal{M}|\max_{\vec{\alpha}\in\mathcal{M}}|f(\vec{\alpha})|\\
&=:
c_{\vec{n},\vec{m}}\me^{-T/4}
+
|\mathcal{M}|\max_{\vec{\alpha}\in\mathcal{M}}|f(\vec{\alpha})|.
\end{split}
\end{equation}
These findings may be summarized as follows. For all $N,n_i,m_i\in\nn$ and every $\epsilon>0$ there is a $c_0>0$ and a compact set $\mathcal{C}\subset\cc^N$ such that
\begin{equation}
\label{gg1}
\left|\int_{\cc^{N}}\md\vec{\alpha}\,f(\vec{\alpha})
\langle\vec{n}|\hat{D}(\vec{\alpha})
|\vec{m}\rangle\right|
\le
\epsilon+c_0
\max_{\vec{\alpha}\in \mathcal{C}}\left|f(\vec{\alpha})\right|
\end{equation}
for all continuous $f:\cc^N\rightarrow\cc$  with $|f(\vec{\alpha})|\le 2$.

Now, for $M\subset\nn^N$ let  
\begin{equation}\label{P}
\hat{P}_M=\sum_{\vec{n}\in M}|\vec{n}\rangle\langle \vec{n}|.
\end{equation} 
This will take the role of a projection onto the relevant part of state space.
For states $\hat{\varrho}_1,\hat{\varrho}_2$ one finds (see the Appendix for a proof)
\begin{equation}
\label{gg2}
\begin{split}
\|\hat{\varrho}_1-\hat{\varrho}_2\|_{\text{tr}}\le
\|\hat{\varrho}_1-\hat{P}_M\hat{\varrho}_1\hat{P}_M+\hat{P}_M\hat{\varrho}_2\hat{P}_M-\hat{\varrho}_2\|_{\text{tr}}\hspace{1cm}\\
+|M|^{3/2}
\max_{\vec{n},\vec{m}  \in M}
\left|\langle\vec{n}|
(\hat{\varrho}_1-\hat{\varrho}_2)|\vec{m}\rangle\right|.
\end{split}
\end{equation}
What is more, using the Heisenberg-Weyl-correspondence,
one has
\begin{equation}
\langle\vec{n}|(\hat{\varrho}_1-\hat{\varrho}_2)|\vec{m}\rangle\propto\int_{\cc^{N}}\md\vec{\alpha}\, \left[\chi_{\hat{\varrho}_1}(\vec{\alpha})-\chi_{\hat{\varrho}_2}(\vec{\alpha})\right]\langle \vec{n}|\hat{D}(\vec{\alpha})|\vec{m}\rangle.
\end{equation}
I.e., combining Eqs.\ (\ref{gg1},\ref{gg2}), one has that for every $N\in\nn$, $\epsilon>0$ and $M\subset\nn^N$ there is a constant $c_0>0$ and 
a compact $\mathcal{C}\subset\cc^N$ such that
\begin{equation}
\begin{split}
\|\hat{\varrho}_1-\hat{\varrho}_2\|_{\text{tr}}&\le
\epsilon+c_0
\max_{\vec{\alpha}\in \mathcal{C}}\left|\chi_{\hat{\varrho}_1}(\vec{\alpha})-\chi_{\hat{\varrho}_2}(\vec{\alpha})\right|\\
&\hspace{2cm}+\|\hat{\varrho}_1-\hat{P}_M\hat{\varrho}_1\hat{P}_M+\hat{P}_M\hat{\varrho}_2\hat{P}_M-\hat{\varrho}_2\|_{\text{tr}}
\end{split}
\end{equation}
for all  $N-$mode states $\hat{\varrho}_1$, $\hat{\varrho}_2$. 

Now, from the gentle measurement lemma of Ref.\ \cite{gentle}
it follows that 
\begin{equation}
\|\hat{\varrho}-\hat{P}_M\hat{\varrho}\hat{P}_M\|_{\text{tr}}\le 2\left({\text{tr}[\hat{\varrho}(\id-\hat{P}_M)]}\right)^{1/2}
\end{equation}
and for $\hat{n}_i=\hat{b}_i^\dagger\hat{b}_i$ one finds (we write $\bar{M}=\nn^N\backslash M$)
\begin{equation}
\begin{split}
N\max_{i}\text{tr}[\hat{\varrho}\hat{n}_i]&
\ge\sum_{i=1}^N\text{tr}[\hat{\varrho}\hat{n}_i]=\sum_{i=1}^N\sum_{\vec{n}\in M}n_i\langle\vec{n}|\hat{\varrho}|\vec{n}\rangle
+
\sum_{i=1}^N\sum_{\vec{n}\in\bar{M}}n_i\langle\vec{n}|\hat{\varrho}|\vec{n}\rangle\\
&\ge
\text{tr}[\hat{\varrho}(\id-\hat{P}_M)]\min_{\vec{n}\in\bar{M}}\sum_{i=1}^Nn_i
>\text{tr}[\hat{\varrho}(\id-\hat{P}_M)]m,
\end{split}
\end{equation} 
where the last inequality holds for the particular choice of 
\begin{equation}
	M=\{\vec{n}\in\nn^N\,:\,\sum_in_i\le m\}.
\end{equation}	
Hence, to summarize, for every $m,N\in\nn$ and $\epsilon>0$  there is a constant $c_0>0$ and 
a compact $\mathcal{C}\subset\cc^N$ such that
\begin{equation}
\begin{split}
\|\hat{\varrho}_1-\hat{\varrho}_2\|_{\text{tr}}&\le
\epsilon+c_0
\max_{\vec{\alpha}\in \mathcal{C}}\left|\chi_{\hat{\varrho}_1}(\vec{\alpha})-\chi_{\hat{\varrho}_2}(\vec{\alpha})\right|\\
&\hspace{2cm}+
2\left({\frac{N}{m}}\right)^{1/2}\left(
\max_i\sqrt{\text{tr}[\hat{\varrho}_1\hat{n}_i]}
+
\max_i\sqrt{\text{tr}[\hat{\varrho}_2\hat{n}_i]}\right)
\end{split}
\end{equation}
for all  $N-$mode states $\hat{\varrho}_1$, $\hat{\varrho}_2$.

\subsection{Trace-norm bounds}

We summarize the above findings in the following Lemma. It shows that closeness
in trace-norm is inherited by pointwise closeness of characteristic functions.
The last line corresponds essentially a bound to the local energy per mode, which can 
in all relevant settings be bounded from above. Note that all constants on the right hand
side of the inequality can be made explicit.

\begin{lemma}[Trace-norm convergence from pointwise phase space convergence]\label{Close}
For every $\epsilon,c_1>0$ and $N\in\nn$ there is a constant $c_0>0$ and a compact $\mathcal{C}\subset\cc^N$ such that
\begin{equation}
\begin{split}
\|\hat{\varrho}_1-\hat{\varrho}_2\|_{\text{tr}}&\le
\epsilon+c_0
\max_{\vec{\alpha}\in \mathcal{C}}\left|\chi_{\hat{\varrho}_1}(\vec{\alpha})-\chi_{\hat{\varrho}_2}(\vec{\alpha})\right|
\end{split}
\end{equation}
for all states $\hat{\varrho}_1$, $\hat{\varrho}_2$ on a collection of $N$ sites with, for all $i$, 
\begin{equation}
\text{tr}[\hat{\varrho}_1\hat{n}_i],\; \text{tr}[\hat{\varrho}_2\hat{n}_i]<c_1.
\end{equation}
\end{lemma}

Using Theorem $\ref{Main}$, the pointwise distance between characteristic functions and $\epsilon$ may hence be subsumed in $\varepsilon$ to yield the trace norm estimate in Corollary $\ref{mainCorr}$.

\section{Summary, discussion and relationship to notions of typicality}

The above theorem shows that weakly clustering states locally relax to maximum 
entropy states. This holds true in the strong sense of trace-norm convergence, to states
that have maximum entropy for given second moments. Physically speaking, this means that locally, the system
appears as if it had entirely relaxed, without the need of an external environment. Every
part of the system forms the environment of the other, and merely mediated by local
interactions, the system relaxes. This is a proven instance of a situation in which 
under non-equilibrium quenched dynamics, closed systems do locally relax. As has been
mentioned before, it can hence be viewed 
as a proof of the ``local relaxation conjecture'', generalizing the findings of Ref.\ \cite{simple},
and constituting to the knowledge of the authors a first proof of local relaxation
in a situation not requiring product or Gaussian, meaning quasi-free, initial states.
That is to say, the statement shows that for a large class of initial states, 
locally, the system quickly looks relaxed and stays like that for long times.

We finally mention a potential link of the above results to the concept of ``typicality'': 
The term typicality refers---roughly speaking---to the idea that when quantum 
states are randomly chosen from a
suitable ensemble of large-dimensional pure states, 
almost all realizations will share properties like reduced states of subsystems
having large entropy, or being close to a Gibbs state. This effect is a manifestation
of the phenomenon of concentration of measure. The simplest such 
situation is realized when drawing random pure
state according to the Haar measure on ${\cal H}=\cc^d\otimes \cc^D$: Then the 
von-Neumann entropy of the $d$-dimensional subsystem will, for large $D$, be almost 
certainly be almost maximum, with any deviations being exponentially suppressed in probability. 
Similarly, if states are drawn from an ensemble of pure states that are eigenvectors of a Hamiltonian
with eigenvalues close to a limiting point in the spectrum, defining an energy,
then one can specify technical conditions under which one can prove that almost certainly, subsystems will
locally look like the reduction of a Gibbs state to that subsystem \cite{Goldstein,Popescu}.
That is to say, almost all states appear to have high local entropy. This is strictly proven only
in case of finite-dimensional constituents, although it is to be expected that the same 
intuition carries over also to the situation of infinite-dimensional constituents. In the light
of this observation, it appears plausible that starting from an ``untypical initial situation'',
systems will eventually be driven to a ``typical one'' by virtue of unitary time evolution.

Needless to say, time evolution defines merely
a one-dimensional manifold, parameterized by time, and from
the above argument alone, there is no guarantee that the system will indeed arrive at a typical
situation (although steps can be taken that show that for most times, albeit unknown ones,
the system appears locally typical \cite{Reimann} in a non-equilibrium situation). 
Here---for a class of models---a significantly stronger statement is proven, in that after a short
time, the system will certainly come arbitrarily close to a maximum entropy state under the constraints
of second moments and will stay there. It would be interesting to take steps to combine
such kinematical and dynamical approaches towards understanding equilibration in 
closed quantum systems.  

\section{Acknowledgements}
We would like to thank T.J.\ Osborne, M.M.\ Wolf, and U.\ Schollw{\"o}ck for discussions.
This work has been supported by the EU (QAP, QESSENCE, COMPAS, MINOS), 
and the EURYI grant scheme.

\section{Appendix: Proofs}

\subsection{Lieb-Robinson bound (Lemma \ref{lemma:lieb_robinson})}
Under periodic boundary conditions, the entries of $C$ take the form $C_{i,j}=C_{i-j}=C_{d_{i,j}}$,
\begin{equation}
C_{l}=\frac{1}{L}\sum_{k=1}^L\me^{2\pi\mi kl/L}\me^{-\mi t\lambda_k},\;\;\;
\lambda_k=-2\cos(2\pi k/L).
\end{equation}
As $A_{i,j}=0$ for $d_{i,j}>1$, we find from Taylor's theorem the estimate
\begin{equation}
\begin{split}
|C_{d_{i,j}}|&=
\left|\frac{1}{L}\sum_{k=1}^L\me^{2\pi\mi kd_{i,j}/L}\sum_{n=d_{i,j}}^\infty\frac{(-\mi t\lambda_k)^n}{n!}\right|
\le
\max_{k\in\mathcal{L}}
\left|\sum_{n=d_{i,j}}^\infty\frac{(-\mi t\lambda_k)^n}{n!}\right|\\
&\le
\max_{k\in\mathcal{L}}
\frac{|t\lambda_k|^{d_{i,j}}}{d_{i,j}!}
\le\frac{|2t|^{d_{i,j}}}{d_{i,j}!}
\le\left(\frac{2\me |t|}{d_{i,j}}\right)^{d_{i,j}},
\end{split}
\end{equation}
i.e., for $\mathcal{A}\subset\mathcal{L}$ we find from Eq.\ (\ref{alpha}) that
\begin{equation}
\begin{split}
\sum_{i\in\mathcal{A}}\left|\alpha_i\right|&\le \sum_{i\in\mathcal{A}} \sum_{j\in \mathcal{S}}|\beta_j||C_{j,i}|
\le\|\vec{\beta}\|_1\max_{j\in \mathcal{S}}\sum_{i\in\mathcal{A}}\left(\frac{2\me |t|}{d_{i,j}}\right)^{d_{i,j}}\\
&\le
\|\vec{\beta}\|_1\max_{j\in \mathcal{S}}\sum_{i\in\mathcal{A}}\left(\frac{2\me |t|}{d_{\mathcal{A},\mathcal{S}}}\right)^{d_{i,j}}\\
&=
\|\vec{\beta}\|_1\max_{j\in \mathcal{S}}\sum_{d=d_{\mathcal{A},\mathcal{S}}}^\infty\left(\frac{2\me |t|}{d_{\mathcal{A},\mathcal{S}}}\right)^{d}\sum_{i\in\mathcal{A}}\delta_{d_{i,j},d}\\
&\le
2
\|\vec{\beta}\|_1\left(\frac{2\me |t|}{d_{\mathcal{A},\mathcal{S}}}\right)^{d_{\mathcal{A},\mathcal{S}}}\sum_{d=0}^\infty\left(\frac{2\me |t|}{d_{\mathcal{A},\mathcal{S}}}\right)^{d}
\le 4
\|\vec{\beta}\|_12^{-d_{\mathcal{A},\mathcal{S}}},
\end{split}
\end{equation}
where the last inequality holds for $4\me |t|\le d_{\mathcal{A},\mathcal{S}}$.

Furthermore, 
\begin{equation}
\begin{split}
C_l&
=
\frac{1}{L}\sum_{k=1}^L\me^{2\mi t\cos(2\pi k/L)}\me^{2\pi \mi kl/L}
=:\frac{1}{L}\sum_{k=1}^Lg(2\pi k/L)\me^{2\pi \mi kl/L},
\end{split}
\end{equation}
where $g(\phi)=\me^{2\mi t\cos(\phi)}$ can be written as
\begin{equation}
\begin{split}
g(\phi)&=\sum_{n=-\infty}^\infty g_n\me^{\mi n\phi},\\
g_n&=\frac{1}{2\pi}\int_0^{2\pi}\md\phi\,g(\phi)
\me^{-\mi n\phi}
=\frac{1}{2\pi}\int_0^{2\pi}\md\phi\,\me^{2\mi t\cos(\phi)}
\me^{\mi n\phi}=\mi^nJ_n(2t)\\
&=\frac{1}{2\pi (\mi n)^2}\int_0^{2\pi}\md\phi\,
\me^{\mi n\phi}\partial_\phi^2\me^{2\mi t\cos(\phi)},
\end{split}
\end{equation}
where $J_n$ is a Bessel function of the first kind. Hence, for $|t|\ge 1$
\begin{equation}
\begin{split}
|g_n|&\le\frac{1}{2\pi n^2}\int_0^{2\pi}\md\phi\,
|\partial_\phi^2\me^{2\mi t\cos(\phi)}|\le
\frac{2|t|+4|t|^2}{ n^2}\le 6\frac{|t|^2}{ n^2}
\end{split}
\end{equation}
and
\begin{equation}
\begin{split}
C_{l}&=\frac{1}{L}\sum_{n=-\infty}^\infty g_n\sum_{k=1}^L\me^{2\pi\mi k(n+l)/L}
=\sum_{n=-\infty}^\infty g_n\delta_{n+l\in L\zz}\\
&=\sum_{z=-\infty}^\infty g_{Lz-l}=\frac{1}{2\pi}\int_0^{2\pi}\md\phi\,g(\phi)
\me^{\mi l\phi}+\sum_{ z=1}^\infty \left(g_{Lz-l}+g_{Lz+l}\right),
\end{split}
\end{equation}
where, for $|t|\ge 1$ and $0\le l\le L/2$, which we can assume as $C_{i,j}=C_{d_{i,j}}$,
\begin{equation}
\begin{split}
\sum_{ z=1}^\infty \left|g_{Lz-l}+g_{Lz+l}\right|&\le \tfrac{6|t|^2}{L^2}\sum_{ z=1}^\infty \left(\tfrac{1}{ (z-l/L)^2}+\tfrac{1}{ (z+l/L)^2}\right)
\le\tfrac{6(\pi^2-4)|t|^2}{L^2},
\end{split}
\end{equation}
i.e., for $L\ge |t|^{7/6}\ge 1$ (we use a bound obtained in Ref.\ \cite{Bessel}),
\begin{equation}
\begin{split}
\left|C_{d_{i,j}}\right|&\le
\left|C_{d_{i,j}}-\frac{1}{2\pi}\int_0^{2\pi}\md\phi\,g(\phi)
\me^{\mi d_{i,j}\phi}\right|+|J_{d_{i,j}}(2t)|\\
&\le
\tfrac{6(\pi^2-4)|t|^2}{L^2}+\tfrac{1}{|2t|^{1/3}}
\le
\tfrac{37}{|t|^{1/3}}.
\end{split}
\end{equation}
\subsection{Moments}
We will need the following bound. 
Let $\epsilon,\mu>0$ and $\mathcal{A}\subset\mathcal{L}$. Then
\begin{equation}
\label{bound_zeta}
\begin{split}
 \sum_{i\in\mathcal{A} }
  \frac{1}{\left[1+d_{i,j}\right]^{2+\mu+\epsilon}}
  &=
  \sum_{r=d_{\mathcal{A},j}}^\infty
  \frac{1}{\left[1+r\right]^{1+\epsilon}\left[1+r\right]^{1+\mu}}
  \sum_{i\in\mathcal{A} }\delta_{d_{i,j},r}\\
&  \le
\frac{2}{\left[1+d_{\mathcal{A},j}\right]^{1+\mu}}
  \sum_{r=0}^\infty
  \frac{1}{\left[1+r\right]^{1+\epsilon}}
  =
\frac{2\zeta(1+\epsilon)}{\left[1+d_{\mathcal{A},j}\right]^{1+\mu}}.
\end{split}
\end{equation}

\subsubsection{Second moments (Lemma \ref{lemma:variances})}
In the following we  assume that $t$ and $L$ are such that $L\ge |t|^{7/6}\ge 1$. Under Assumption \ref{assumption:twoPoint}
we find  from Eqs.\ (\ref{moments},\ref{bound_zeta}) and Lemma \ref{lemma:lieb_robinson} that
\begin{equation}
\begin{split}
|\sigma_{\mathcal{A},\mathcal{B}}|&\le
 \sum_{\substack{i\in\mathcal{A} \\ j\in\mathcal{B}}}
 |\alpha_i\alpha_j| \Bigl(
\left|
\langle\hat{b}_i^\dagger\hat{b}_j\rangle\right|
+\left|
\langle\hat{b}_j\hat{b}_i^\dagger\rangle\right|
\Bigr)\\
&\le
\frac{37^2}{|t|^{2/3}}\|\vec{\beta}\|^2_1 \sum_{\substack{i\in\mathcal{A} \\ j\in\mathcal{B}}}
\Bigl(
2\left|
\langle\hat{b}_i^\dagger\hat{b}_j\rangle\right|
+\delta_{i,j}\Bigr)\\
&\le
\frac{37^2(2\cSecondMomentsAmp+1)}{|t|^{2/3}} \|\vec{\beta}\|^2_1 \sum_{\substack{i\in\mathcal{A} \\ j\in\mathcal{B}}}
\frac{1}{\left[1+d_{i,j}\right]^{2+\cSecondMomentsMu+\cSecondMomentsEps}}\\
&\le
37^2(4\cSecondMomentsAmp+2)\zeta(1+\cSecondMomentsEps)\|\vec{\beta}\|^2_1  
\frac{\min\{|\mathcal{A}|,|\mathcal{B}|\}}{|t|^{2/3}\left[1+d_{\mathcal{A},\mathcal{B}}\right]^{1+\cSecondMomentsMu}}
\end{split}
\end{equation}
for all $\mathcal{A},\mathcal{B}\subset\mathcal{L}$. 
Similarly, for all $\mathcal{A}_i\subset\mathcal{L}$, $i=1,\dots,n$, with $\mathcal{A}_i\cap \mathcal{A}_j=\emptyset$ for $i\ne j$, we find ($\mathcal{A}:=\bigcup_i\mathcal{A}_i$)
\begin{equation}
\begin{split}
\sum_{\substack{i,j=1\\ i\ne j}}^n|\sigma_{\mathcal{A}_i,\mathcal{A}_j}|&\le
\tfrac{37^2(2\cSecondMomentsAmp+1)}{|t|^{2/3}} \|\vec{\beta}\|^2_1\sum_{i=1}^n
 \sum_{k\in\mathcal{A}_i }
 \sum_{ l\in\mathcal{A}\backslash\mathcal{A}_i }
\tfrac{1}{\left[1+d_{k,l}\right]^{2+\cSecondMomentsMu+\cSecondMomentsEps}}\\
&\le
\cVariancesFP \|\vec{\beta}\|^2_1\sum_{i=1}^n
 \tfrac{|\mathcal{A}_i|}{|t|^{2/3}\left[1+d_{\mathcal{A}_i,\mathcal{A}\backslash\mathcal{A}_i}\right]^{1+\cSecondMomentsMu}}
.
\end{split}
\end{equation}
Now, for all $\mathcal{A},\mathcal{B}\subset\mathcal{L}$ with $\mathcal{A}\cap\mathcal{B}=\emptyset$ one has $\sigma_{\mathcal{A}\cup\mathcal{B}}=\sigma_{\mathcal{A}}+\sigma_{\mathcal{B}}+2\sigma_{\mathcal{A},\mathcal{B}}$, 
i.e.,
\begin{equation}
\begin{split}
\sigma_{\mathcal{L}}&=
\sigma_{\mathcal{L}\backslash\mathcal{A}}+2\sigma_{\mathcal{L}\backslash\mathcal{A},\mathcal{A}}+\sigma_{\mathcal{A}}
=
\sigma_{\mathcal{L}\backslash\mathcal{A}}+2\sigma_{\mathcal{L}\backslash\mathcal{A},\mathcal{A}}+
\sum_{i=1}^n\sigma_{\mathcal{A}_i}+\sum_{\substack{i,j=1\\ i\ne j}}^n\sigma_{\mathcal{A}_i,\mathcal{A}_j},
\end{split}
\end{equation}
i.e.,
\begin{equation}
\begin{split}
\left|\sigma_{\mathcal{L}}-\sum_{i=1}^n\sigma_{\mathcal{A}_i}\right|&\le|\sigma_{\mathcal{L}\backslash\mathcal{A}}|+2|\sigma_{\mathcal{L}\backslash\mathcal{A},\mathcal{A}}|
+\cVariancesFP\|\vec{\beta}\|^2_1 \sum_{i=1}^n
\tfrac{|\mathcal{A}_i|}{|t|^{2/3}\left[1+d_{\mathcal{A}\backslash \mathcal{A}_i,\mathcal{A}_i}\right]^{1+\cSecondMomentsMu}}.
\end{split}
\end{equation}

Finally, under Assumption \ref{assumption:twoPoint}
 and for $t$, $\mathcal{A}$ such that $4\me\le 4\me |t|\le d_{\mathcal{A},\mathcal{S}}$, we find from Eqs.\ (\ref{moments},\ref{bound_zeta}) and Lemma \ref{lemma:lieb_robinson} that
\begin{equation}
\begin{split}
|\sigma_{\mathcal{A},\mathcal{B}}|&\le 
 \sum_{\substack{i\in\mathcal{A} \\ j\in\mathcal{B}}}
\frac{ (2\cSecondMomentsAmp+1)|\alpha_i\alpha_j|}{\left[1+d_{i,j}\right]^{1+\cSecondMomentsMu+\cSecondMomentsEps}}
\le
 \sum_{i\in\mathcal{A}}|\alpha_i|
\max_{i\in\mathcal{A}}\sum_{j\in\mathcal{B}}
\frac{37(2\cSecondMomentsAmp+1)\|\vec{\beta}\|_1}{\left[1+d_{i,j}\right]^{1+\cSecondMomentsMu+\cSecondMomentsEps}}\\
&\le
37(4\cSecondMomentsAmp+2)\|\vec{\beta}\|_1\zeta(1+\cSecondMomentsEps) \sum_{i\in\mathcal{A}}|\alpha_i|
\le \cVariancesFP\|\vec{\beta}\|^2_12^{-d_{\mathcal{A},\mathcal{S}}}.
\end{split}
\end{equation}



\subsubsection{Fourth moments (Lemma \ref{lemma:fourthmoments})}
From Eq.\ (\ref{moments}) we find
\begin{equation}
\begin{split}
|f_{\mathcal{A}}|
&\le
\sum_{i,j,k,l\in\mathcal{A}}|\alpha_i\alpha_j\alpha_k\alpha_l|
\Bigl(|\langle\hat{b}_i^\dagger\hat{b}_j^\dagger\hat{b}_k
\hat{b}_l \rangle|+|\langle\hat{b}_i^\dagger\hat{b}_k\hat{b}_j^\dagger\hat{b}_l \rangle|+|\langle\hat{b}_i^\dagger\hat{b}_k\hat{b}_l\hat{b}_j^\dagger \rangle|
\\
&\hspace{4cm}+|\langle\hat{b}_k\hat{b}_j^\dagger\hat{b}_i^\dagger\hat{b}_l \rangle|+|\langle\hat{b}_k\hat{b}_j^\dagger\hat{b}_l\hat{b}_i^\dagger \rangle|
+|\langle\hat{b}_k\hat{b}_l\hat{b}_i^\dagger
\hat{b}_j^\dagger \rangle|
\Bigr)\\
&\le
3\sum_{i,j,k,l\in\mathcal{A}}|\alpha_i\alpha_j\alpha_k\alpha_l|
\Bigl(2|\langle\hat{b}_i^\dagger\hat{b}_j^\dagger\hat{b}_k
\hat{b}_l \rangle|+4\delta_{k,j}|\langle\hat{b}_i^\dagger\hat{b}_l \rangle|
+\delta_{k,j}\delta_{i,l}
\Bigr),
 \end{split}
\end{equation}
where, under Assumption \ref{assumption:twoPoint}, we have 
\begin{equation}
\begin{split}
\left(
\delta_{i,j}+
4|\langle\hat{b}_i^\dagger\hat{b}_j\rangle|
\right)\delta_{k,l}\le
\tfrac{(4\cSecondMomentsAmp+1) \delta_{k,l}}{\left[1+d_{i,j}\right]^{1+\cSecondMomentsEps}}
\le
\tfrac{(4\cSecondMomentsAmp+1)}{\left([1+d_{i,j}][1+d_{k,l}]\right)^{1+\cSecondMomentsEps}}\\
\le
\sum_{(r,s,t,u)\in P(i,j,k,l)}\tfrac{(4\cSecondMomentsAmp+1)}{\left([1+d_{r,s}][1+d_{t,u}]\right)^{1+\cSecondMomentsEps}},
  \end{split}
\end{equation}
i.e., under Assumptions \ref{assumption:twoPoint} and \ref{assumption:fourPoint} we have 
\begin{equation}
\begin{split}
|f_{\mathcal{A}}|
&\le
\sum_{i,j,k,l\in\mathcal{A}}\!\!\!\!|\alpha_i\alpha_j\alpha_k\alpha_l|\sum_{(r,s,t,u)\in P(i,j,k,l)}\tfrac{6\cFourthMomentsAmp+3(4\cSecondMomentsAmp+1)}{\left([1+d_{r,s}][1+d_{t,u}]\right)^{1+\min\{\cSecondMomentsEps,\cFourthMomentsEps\}}}.
  \end{split}
\end{equation}
Now let $(r,s,t,u)$ be a given permutation of $(i,j,k,l)$. Then, using the geometric mean inequality and Eq.\ (\ref{bound_zeta}),
\begin{equation}
\begin{split}
\sum_{i,j,k,l\in\mathcal{A}}
\tfrac{ |\alpha_i\alpha_j| |\alpha_k\alpha_l|}{\left([1+d_{r,s}][1+d_{t,u}]\right)^{1+\min\{\cSecondMomentsEps,\cFourthMomentsEps\}}}
&=
\sum_{r,s,t,u\in\mathcal{A}}
\tfrac{ |\alpha_r||\alpha_s| |\alpha_t||\alpha_u|}{\left([1+d_{r,s}][1+d_{t,u}]\right)^{1+\min\{\cSecondMomentsEps,\cFourthMomentsEps\}}}\\
&=\Bigl[
\sum_{r,s\in\mathcal{A}}
\tfrac{ |\alpha_r||\alpha_s|}{[1+d_{r,s}]^{1+\min\{\cSecondMomentsEps,\cFourthMomentsEps\}}}\Bigr]^2\\
&\le
\Bigl[
\sum_{r,s\in\mathcal{A}}
\tfrac{ |\alpha_r|^2}{[1+d_{r,s}]^{1+\min\{\cSecondMomentsEps,\cFourthMomentsEps\}}}\Bigr]^2
\\
&\le
4\zeta^2(1+\min\{\cSecondMomentsEps,\cFourthMomentsEps\})\Bigl[\sum_{i\in\mathcal{A}}|\alpha_i|^2\Bigr]^2,
  \end{split}
\end{equation}
i.e., using Lemma \ref{lemma:lieb_robinson} and the fact that $\sum_{i\in\mathcal{L}}|\alpha_i|^2=\|\vec{\beta}\|_2^2$ and $|P|=24$, we find
for $L\ge |t|^{7/6}\ge 1$ 
\begin{equation}
\begin{split}
\sum_{i=1}^n|f_{\mathcal{A}_i}|
&\le
96(6\cFourthMomentsAmp+3(4\cSecondMomentsAmp+1))\zeta^2(1+\min\{\cSecondMomentsEps,\cFourthMomentsEps\})\sum_{i=1}^n\Bigl[\sum_{j\in\mathcal{A}_i}|\alpha_j|^2\Bigr]^2\\
&\le  
96(6\cFourthMomentsAmp+3(4\cSecondMomentsAmp+1))\zeta^2(1+\min\{\cSecondMomentsEps,\cFourthMomentsEps\})
\|\vec{\beta}\|_2^2
\max_{i}\sum_{j\in\mathcal{A}_i}|\alpha_j|^2\\
&\le  
96(6\cFourthMomentsAmp+3(4\cSecondMomentsAmp+1))37^2\zeta^2(1+\min\{\cSecondMomentsEps,\cFourthMomentsEps\})
\|\vec{\beta}\|_2^2
\|\vec{\beta}\|^2_1
\tfrac{\max_{i}|\mathcal{A}_i|}{|t|^{2/3}}.
  \end{split}
\end{equation}

\subsection{Blocking argument (Lemma \ref{lemma:bernstein})}
In the following we write $\tau=8\me |t|$.
\subsubsection{Distances}
For all $k\in A_i$, $l\in A_j$ we have (we pick $s\in\mathcal{S}$ such that $d_{l,s}=d_{l,\mathcal{S}}$)
\begin{equation}
(i-1)(a+b)\le d_{k,\mathcal{S}}\le d_{k,s}\le d_{k,l}+d_{l,s}=d_{k,l}+d_{l,\mathcal{S}}< d_{k,l}+j(a+b)-b
\end{equation}
and (we pick $s\in\mathcal{S}$ such that $d_{k,s}=d_{k,\mathcal{S}}$)
\begin{equation}
(j-1)(a+b)\le d_{l,\mathcal{S}}\le d_{l,s}\le d_{l,k}+d_{k,s}=d_{l,k}+d_{k,\mathcal{S}}< d_{k,l}+i(a+b)-b,
\end{equation}
i.e.,
\begin{equation}
d_{\mathcal{A}_i,\mathcal{A}_j}=\min_{\substack{k\in \mathcal{A}_i\\ l\in \mathcal{A}_j}}d_{k,l}>
|i-j|(a+b)-a.
\end{equation}
Furthermore
\begin{equation}
d_{\mathcal{S},\mathcal{T}}=\min_{ l\in \mathcal{T}}d_{\mathcal{S},l}\ge n(a+b)\ge \frac{\tau}{2}.
\end{equation}

\subsubsection{Cardinalities}
For $i\in\mathcal{L}\backslash\mathcal{S}$ pick $s_i\in\partial\mathcal{S}$ such that $d_{i,\mathcal{S}}=d_{i,s_i}=d_{i,\partial\mathcal{S}}$. Then
\begin{equation}
\begin{split}
|\mathcal{A}_1|&=\sum_{0\le l< a}\sum_{i\in\mathcal{L}}\delta_{d_{i,\mathcal{S}},l}=|\mathcal{S}|+\sum_{1\le l< a}\sum_{i\in\mathcal{L}\backslash \mathcal{S}}\delta_{d_{i,\mathcal{S}},l}=|\mathcal{S}|+\sum_{1\le l< a}\sum_{i\in\mathcal{L}\backslash \mathcal{S}}\delta_{d_{i,s_i},l}\\
&\le
|\mathcal{S}|+\sum_{1\le l< a}\sum_{s\in\partial\mathcal{S}}\sum_{i\in\mathcal{L}\backslash \mathcal{S}}\delta_{d_{i,s},l}
\le
|\mathcal{S}|+2|\partial\mathcal{S}|a.
\end{split}
\end{equation}
Similarly, for $i\ne 1$
\begin{equation}
\begin{split}
|\mathcal{A}_i|&\le
2|\partial\mathcal{S}|\sum_{(i-1)(a+b)\le l< i(a+b)-b}1\le
2|\partial\mathcal{S}|a
\end{split}
\end{equation}
i.e.,
\begin{equation}
\begin{split}
|\mathcal{A}|=\sum_{i=1}^n|\mathcal{A}_i|\le|\mathcal{S}|+2|\partial\mathcal{S}|na\le
|\mathcal{S}|+2|\partial\mathcal{S}|\tau.
\end{split}
\end{equation}
Furthermore,
\begin{equation}
\begin{split}
|\mathcal{B}|&\le
2|\partial\mathcal{S}|\sum_{i=1}^n\sum_{i(a+b)-b\le l< i(a+b)}1
\le
2|\partial\mathcal{S}|nb\le
2|\partial\mathcal{S}|\frac{\tau b}{a}.
\end{split}
\end{equation}

\subsubsection{Moments}
In the following let $t$ and $L$ be such that
\begin{equation}
L^{6/7}\ge |t|\ge 2,\;\;\;
\log |t|\le |t|^{1/3+\mu},
\end{equation}
where $\mu=\cSecondMomentsMu/(6(\cSecondMomentsMu+1))$.
Then 
\begin{equation}
\frac{|t|^{2/3}}{\log |t|}=a\ge b=|t|^{1/3-\mu}\ge 1,\;\;\; \left\lfloor\frac{8\me |t|}{a+b}\right\rfloor =n>1.
\end{equation}
Combining the above bounds on distances and cardinalities with Lemmas \ref{lemma:variances}, \ref{lemma:fourthmoments} we find under Assumptions \ref{assumption:twoPoint},\ref{assumption:fourPoint} that
\begin{equation}
\begin{split}
|\sigma_{\mathcal{T}}|,|\sigma_{\mathcal{B},\mathcal{T}}|,|\sigma_{\mathcal{A},\mathcal{T}}|&\le 
\cVariancesFP
\|\vec{\beta}\|^2_12^{-4\me |t|},\\
\end{split}
\end{equation}
and
\begin{equation}
\begin{split}
|\sigma_{\mathcal{B}}|,|\sigma_{\mathcal{A},\mathcal{B}}|&\le 
2\cVariancesFP |\partial\mathcal{S}| \|\vec{\beta}\|^2_1  
\frac{\tau b}{a|t|^{2/3}}
=
16\me\cVariancesFP |\partial\mathcal{S}| \|\vec{\beta}\|^2_1  
\frac{ |t|^{1/3} b}{a}\\
&=16\me\cVariancesFP |\partial\mathcal{S}| \|\vec{\beta}\|^2_1  
   |t|^{-\mu}\log|t|
,\\
\left|\sigma_{\mathcal{L}}-\sum_{i=1}^n\sigma_{\mathcal{A}_i}\right|&\le
|\sigma_{\mathcal{B}\cup\mathcal{T}}|+2|\sigma_{\mathcal{B}\cup\mathcal{T},\mathcal{A}}|
+\cVariancesFP\|\vec{\beta}\|^2_1
\tfrac{|\mathcal{A}|}{|t|^{2/3}b^{1+\cSecondMomentsMu}}\\
&\le
|\sigma_{\mathcal{B}}|+|\sigma_{\mathcal{T}}|+2|\sigma_{\mathcal{B},\mathcal{T}}|+2|\sigma_{\mathcal{B},\mathcal{A}}|
+2|\sigma_{\mathcal{A},\mathcal{T}}|\\
&\hspace{2cm}+\cVariancesFP\|\vec{\beta}\|^2_1\left(
\tfrac{|\mathcal{S}|}{|t|^{2/3}}
+16\me|\partial\mathcal{S}|\tfrac{ |t|^{1/3}}{b^{1+\cSecondMomentsMu}}\right)\\
&\le
5\cVariancesFP
\|\vec{\beta}\|_1^22^{-4\me |t|}
\\
&\hspace{1cm}+\cVariancesFP\|\vec{\beta}\|^2_1\left(
\tfrac{|\mathcal{S}|}{|t|^{2/3}}
+16\me|\partial\mathcal{S}| \left[3|t|^{-\mu}\log |t|+\frac{|t|^{1/3}}{b^{1+\cSecondMomentsMu}}\right]\right)\\
&\le
\cVariancesFP\|\vec{\beta}\|^2_1\left(
\tfrac{2|\mathcal{S}|}{|t|^{2/3}}
+80\me|\partial\mathcal{S}| |t|^{-\mu}\log |t|\right)
,\\
\sum_{i=1}^n|f_{\mathcal{A}_i}|
&\le
\cFP
\|\vec{\beta}\|_2^2\|\vec{\beta}\|_1^2
\frac{|\mathcal{A}_1|}{|t|^{2/3}}
\le
\cFP
\|\vec{\beta}\|_2^2\|\vec{\beta}\|_1^2
\tfrac{|\mathcal{S}|+2|\partial\mathcal{S}|a}{|t|^{2/3}}\\
&=\cFP
\|\vec{\beta}\|_2^2\|\vec{\beta}\|_1^2\left(
\tfrac{|\mathcal{S}|}{|t|^{2/3}}+\tfrac{2|\partial\mathcal{S}|}{\log |t|}\right).
  \end{split}
\end{equation}

\subsection{Closeness of quantum states}

We have for any states $\hat{\varrho}_1,\hat{\varrho}_2$,
\begin{equation}
\begin{split}
\|\hat{\varrho}_1-\hat{\varrho}_2\|_{\text{tr}}\le
\|\hat{\varrho}_1-\hat{P}_M\hat{\varrho}_1\hat{P}_M+\hat{P}_M\hat{\varrho}_2\hat{P}_M-\hat{\varrho}_2\|_{\text{tr}}\\
+\|\hat{P}_M\hat{\varrho}_1\hat{P}_M-\hat{P}_M\hat{\varrho}_2\hat{P}_M\|_{\text{tr}},
\end{split}
\end{equation}
treating the relevant part of state space. 
Within that relevant part, we can make
use of
\begin{equation}
\begin{split}
\|\hat{P}_M\hat{\varrho}_1\hat{P}_M-\hat{P}_M\hat{\varrho}_2\hat{P}_M\|_{\text{tr}}&=\|\hat{P}_M(\hat{\varrho}_1-\hat{\varrho}_2)\hat{P}_M\|_{\text{tr}} \\
&=
\text{tr}\left[\left({\hat{P}_M(\hat{\varrho}_1-\hat{\varrho}_2)\hat{P}_M(\hat{\varrho}_1-\hat{\varrho}_2)\hat{P}_M}\right)^{1/2}\right]\\
&=
\sum_{\vec{n}}\langle\vec{n}|
\left({\hat{P}_M(\hat{\varrho}_1-\hat{\varrho}_2)\hat{P}_M(\hat{\varrho}_1-\hat{\varrho}_2)\hat{P}_M}\right)^{1/2}
|\vec{n}\rangle,
\end{split}
\end{equation}
giving rise to the bound
\begin{equation}
\begin{split}
\|\hat{P}_M\hat{\varrho}_1\hat{P}_M-\hat{P}_M\hat{\varrho}_2\hat{P}_M\|_{\text{tr}}&\le
\sum_{\vec{n}}
\left({\langle\vec{n}|
\hat{P}_M(\hat{\varrho}_1-\hat{\varrho}_2)\hat{P}_M(\hat{\varrho}_1-\hat{\varrho}_2)\hat{P}_N
|\vec{n}\rangle}\right)^{1/2}
\\
&=
\sum_{\vec{n}}
\left({
\sum_{\vec{m}}
\left|\langle\vec{n}|
\hat{P}_M(\hat{\varrho}_1-\hat{\varrho}_2)\hat{P}_M|\vec{m}\rangle\right|^2}\right)^{1/2}
\\
&=
\sum_{\vec{n} \in M}
\left({
\sum_{\vec{m} \in M}
\left|\langle\vec{n}|
(\hat{\varrho}_1-\hat{\varrho}_2)|\vec{m}\rangle\right|^2}\right)^{1/2}\\
&\le
|M|^{3/2}
\max_{\vec{n},\vec{m}  \in M}
\left|\langle\vec{n}|
(\hat{\varrho}_1-\hat{\varrho}_2)|\vec{m}\rangle\right| .
\end{split}
\end{equation}


\begin{thebibliography}{99}

\bibitem{Robinson}
	O.\ E.\ Lanford, III, and D.\ W.\ Robinson, 
	Commun.\ Math.\ Phys.\
	{\bf 24}, 193 (1972).
	
\bibitem{Spohn}
	T.\ V.\ Dudnikova, A.\ I.\ Komech, and H.\ Spohn, 
	J.\ Math.\ Phys.\ {\bf 44}, 2596 (2003).

\bibitem{Spohn2}
	T.\ V.\ Dudnikova and H.\ Spohn, 
	Markov Processes and Related Fields {\bf 12}, 645 (2006).
	
\bibitem{Lieb}
	B.\ M.\ McCoy and E.\ Barouch, Phys.\ Rev.\ A {\bf 3}, 786
	(1971).
	
\bibitem{Tegmark}
	M.\ Tegmark and L.\ Yeh, Physica A {\bf 202}, 342 (1994).
	
\bibitem{Barthel}
	T.\ Barthel and U.\ Schollw{\"o}ck, Phys.\ Rev.\ Lett.\ {\bf 100}, 
	100601 (2008).
	
\bibitem{simple}
	M.\ Cramer, C.M.\ Dawson, J.\ Eisert, and T.\ J.\ Osborne,
	Phys.\ Rev.\ Lett.\ {\bf 100}, 030602 (2008).

\bibitem{Calabrese}
    	P.\ Calabrese and J.\ Cardy, Phys.\ Rev.\ Lett.\ {\bf 96},
    	136801 (2006).

\bibitem{Sachdev}
	K.\ Sengupta, S.\ Powell, and S.\ Sachdev,
    	Phys.\ Rev.\ A {\bf 69}, 053616 (2004).
	
\bibitem{Kehrein}
	M.\ M{\"o}ckel and S.\ Kehrein, 
	Phys.\ Rev.\ Lett.\ {\bf 100}, 175702
 	(2008).	
    
\bibitem{Kollath}
	C.\ Kollath, A.\ M.\ L{\"a}uchli, and E.\ Altman, 
	Phys.\ Rev.\ Lett.\ {\bf 98}, 180601 (2007).
    
\bibitem{Flesch} 
    	A.\ Flesch, M.\ Cramer,
	I.\ P.\ McCulloch, U.\ Schollw{\"o}ck, 
	and J.\ Eisert, Phys.\ Rev.\ A {\bf 78}, 033608 (2008).
	
\bibitem{Demler}
	P.\ Barmettler, A.\ M.\ Rey, E.\ A.\ Demler, M.\ D.\ Lukin, I.\  Bloch,
	and V.\ Gritsev, 
	Phys.\ Rev.\ A {\bf 78}, 012330 (2008).

\bibitem{RigolNew}
	M.\ Rigol, V.\ Dunjko, V.\ Yurovsky, and M.\ Olshanii, 
	Phys.\ Rev.\ Lett.\ {\bf 98}, 050405 (2007).
	
\bibitem{general} 
	In preparation (2010).
	
\bibitem{QCA}
	J.\ Guetschow, S.\ Uphoff, R.\ F.\ Werner, and Z.\ Zimboras,
	J.\ Math.\ Phys.\ {\bf 51}, 015203 (2010).
	
\bibitem{Goldstein}
	S.\ Goldstein, J.L.\ Lebowitz, R.\ Tumulka, and N.\ Zanghi, 
	Phys.\ Rev.\ Lett.\ {\bf 96}, 050403 (2006).
	
\bibitem{Popescu}
	S.\ Popescu, A.J.\ Short, and A.\ Winter, 
	Nature Physics {\bf 2}, 754 (2006).	

\bibitem{Reimann}	
	P.\ Reimann, 
	Phys.\ Rev.\ Lett.\ {\bf 101}, 190403 (2008).

\bibitem{LR}	
	E.\ H.\ Lieb and D.\ W.\ Robinson, 
	Commun.\ Math.\ Phys.\ {\bf 28}, 251 (1972).

\bibitem{LR2}
	B.\ Nachtergaele, Y.\ Ogata, and R.\ Sims, 
	J.\ Stat.\ Phys.\ {\bf 124}, 1 (2006).
	
\bibitem{LRH}		
	M.\ B.\ Hastings and T.\ Koma, Comm.\
	Math.\ Phys.\  {\bf 265}, 781 (2006).

\bibitem{Area}
	J.\ Eisert, M.\ Cramer, and M.B.\ Plenio, 
	Rev.\ Mod.\ Phys. {\bf 82}, 277 (2010).
	
\bibitem{LR3}
	M.\ Cramer, A.\ Serafini, and J.\ Eisert, in
	{\it Quantum information and many body quantum systems}, Eds.\ M. Ericsson, S. Montangero, 
	Pisa: Edizioni della Normale, pp 51-72, 2008 (Publications of the Scuola Normale Superiore. CRM Series, 8).

\bibitem{LR4}
	B.\ Nachtergaele, H.\ Raz, B.\ Schlein, and R.\ Sims,
	arXiv:0712.3820. 

\bibitem{Wolf}
	M.\ M.\ Wolf, G.\ Giedke, and J.\ I.\ Cirac,
	Phys.\ Rev.\ Lett.\ {\bf 96}, 080502 (2006).

\bibitem{hudson} 
	C.\ D.\ Cushen and R.\ L.\ Hudson, 
	J.\ Appl.\ Prob.\ {\bf 8}, 454 (1971).
    
\bibitem{trace}
    	E.\ B.\ Davies, Commun.\ Math.\ Phys.\ {\bf 15},
    	277 (1969); {\bf 27}, 309 (1972).

\bibitem{gentle}
	A.\ Winter, 
	IEEE Trans.\ Inf.\ Theory {\bf 45}, 2481 (1999).
		
\bibitem{Bessel} 
	L.\ Landau, Electron. J.\ Diff.\ Eqns., Conf.\ {\bf 04}, 
	147 (2000).
	
	
	
\end{thebibliography}
\end{document}